%% file: main.tex
\documentclass[10pt,conference]{IEEEtran}

\usepackage{algorithm}
\usepackage{algpseudocode}
\usepackage{amssymb}
\usepackage{xspace}
\usepackage{makecell}
\usepackage{float}
\usepackage{hyperref}
\usepackage{subcaption}
\usepackage{paralist}
\usepackage{graphicx}
\usepackage{mdframed}
\usepackage[inline]{enumitem}
\usepackage{import}
\usepackage{pdfpages}

\DeclareMathAlphabet{\mathcal}{OMS}{cmsy}{m}{n}

\mdfdefinestyle{box}{%
	linecolor=black,
	outerlinewidth=2pt,
	innertopmargin=4pt,
	innerbottommargin=4pt,
	innerrightmargin=4pt,
	innerleftmargin=4pt,
	leftmargin = 4pt,
	rightmargin = 4pt
}

\makeatletter
\def\namedlabel#1#2{\begingroup
	#2%
	\def\@currentlabel{#2}%
	\phantomsection\label{#1}\endgroup
}
\makeatother

\algrenewcommand\algorithmicprocedure{}
\algtext*{EndProcedure}
\algtext*{EndWhile}
\algtext*{EndFor}
\algtext*{EndIf}
\algblockdefx[NAME]{START}{END}%
[3][Unknown]{#2 #1#3}%
{}
\algtext*{END}

\algnewcommand\And{\textbf{ and }}
\algnewcommand\Or{\textbf{ or }}

\newcommand{\thesystem}{\textsc{ezBFT}\xspace}

\begin{document}

\title{\thesystem: Decentralizing Byzantine Fault-Tolerant State Machine Replication}

\author{\IEEEauthorblockN{Balaji Arun}
\IEEEauthorblockA{\textit{Virginia Tech} \\
Blacksburg, VA, USA \\
balajia@vt.edu}
\and
\IEEEauthorblockN{Sebastiano Peluso}
\IEEEauthorblockA{\textit{Amazon Web Services} \\
Seattle, WA, USA \\
peluso.sebastiano@gmail.com}
\and
\IEEEauthorblockN{Binoy Ravindran}
\IEEEauthorblockA{\textit{Virginia Tech} \\
Blacksburg, VA, USA\\
binoy@vt.edu}
}

\maketitle

\begin{abstract}
We present \thesystem, a novel leaderless, distributed consensus protocol capable of tolerating byzantine faults. \thesystem's main goal is to minimize the client-side latency in WAN deployments. It achieves this by (i) having no designated primary replica, and instead, enabling every replica to order the requests that it receives from clients; (ii) using only three communication steps to order requests in the common case; and (iii) involving clients actively in the consensus process. In addition, \thesystem minimizes the potentially negative effect
of a byzantine replica on the overall system performance. We developed \thesystem's formal specification in TLA+, show that it provides the classic properties of BFT protocols including consistency, stability, and liveness, and developed an implementation. Our experimental evaluation reveals that  \thesystem improves client-side latency by as much as 40\% over state-of-the-art byzantine fault-tolerant protocols including PBFT, FaB, and Zyzzyva.
\end{abstract}


\input{intro.tex}
\input{model.tex}

\input{overview.tex}

\input{design.tex}

\input{correctness.tex}

\input{evaluation.tex}

\input{relwork.tex}

\input{conclusion.tex}

\bibliographystyle{IEEEtran}
\bibliography{ezbft}

\newpage
\onecolumn

\appendix

\subsection{TLA+ Specification}



\section*{Supplementary material}

APPENDIX

*A. TLA+ Specification*

$TLC$ Model Values: Client ← [model value] < symmetrical > $\{c0\}$
$FakeReplica$ ← [model value] < symmetrical > $\{r0\}$
$GoodReplica$ ← [model value] < symmetrical > $\{r1, r2, r3\}$ $Commands$ ← [model value] < symmetrical > $\{cmd0\}$
$FastQuorum(r)$ ← {CASE $r = r0 \to \{r0, r1, r2, r3\}$
 □ $r = r1 \to \{r1, r2, r3, r0\}$
 □ $r = r2 \to \{r1, r2, r3, r0\}$
 □ $r = r3 \to \{r1, r2, r3, r0\}$}
$SlowQuourum(r)$ ← {CASE $r = r0 \to \{r0, r1, r2\}$
 □ $r = r1 \to \{r1, r2, r3\}$
 □ $r = r2 \to \{r1, r2, r3\}$
 □ $r = r3 \to \{r1, r2, r3\}$}
$MaxOwners$ ← 1
$NextOwner(r)$ ← CASE $r = r0 \to r1$
 □ $r = r1 \to r2$
 □ $r = r2 \to r3$
 □ $r = r3 \to r0$
$WeakQuorum(r)$ ← {CASE $r = r0 \to \{r0, r1\}$
 □ $r = r1 \to \{r1, r2\}$
 □ $r = r2 \to \{r1, r2\}$
 □ $r = r3 \to \{r1, r3\}$}

───────────── MODULE $ezBFT$ ─────────────

EXTENDS $Integers$, $FiniteSets$, $TLC$, $Sequences$

────────────────────────────────────────

$Max(S) \triangleq$ IF $S = \{\}$ THEN 0 ELSE CHOOSE $i \in S : \forall j \in S : j \leq i$

$PT \triangleq$ INSTANCE $PT$

Constant parameters: Commands: the set of all possible commands Clients: the set of all possible commands $GoodReplica$: the set of all possible commands $FakeReplicas$: the set of all $ezBFT$ replicas
  $FastQuorums(r)$: the set of all fast quorums where $r$ is a command leader
  $SlowQuorums(r)$: the set of all slow quorums where $r$ is a command leader Commands: the set of all possible commands

CONSTANTS $Commands$, $Clients$, $GoodReplicas$, $FakeReplicas$, $FastQuorums(\_)$,
         $SlowQuorums(\_)$, $WeakQuorums(\_)$, $MaxOwners$, $NextOwner(\_)$

$Replicas \triangleq GoodReplicas \cup FakeReplicas$

ASSUME $IsFiniteSet(Replicas)$

ASSUME $Cardinality(Replicas) = (3 * Cardinality(FakeReplicas)) + 1$

Quorum conditions: (simplified)

ASSUME $\forall\, r \in Replicas$ :
  $\wedge SlowQuorums(r) \subseteq$ SUBSET $Replicas$

1

$\quad\quad \wedge\, \forall\, SQ \in SlowQuorums(r):$
$\quad\quad\quad \wedge\, r \in SQ$
$\quad\quad\quad \wedge\, Cardinality(SQ) = ((Cardinality(Replicas) * 2) + 1) \div 3$

ASSUME $\forall\, r \in Replicas:$
$\quad \wedge\, FastQuorums(r) \subseteq$ SUBSET $Replicas$
$\quad \wedge\, \forall\, FQ \in FastQuorums(r):$
$\quad\quad \wedge\, r \in FQ$
$\quad\quad \wedge\, Cardinality(FQ) = Cardinality(Replicas)$

Special none command

$none\ \triangleq\ $ CHOOSE $c : c \notin Commands$

The instance space

$Instances\ \triangleq\ Replicas \times (1\mathrel{{.}\,{.}} Cardinality(Commands))$

The possible status of a command in the log of a replica.

$Status\ \triangleq\ \{\,\text{``not-seen''},\ \text{``spec-ordered''},\ \text{``committed''}\,\}$

All possible protocol messages:

$SpecOrderMessage\ \triangleq$
$\quad\quad [type : \{\,\text{``spec-order''}\,\},\ src : Replicas,\ dst : Replicas,$
$\quad\quad\ inst : Instances,\ owner : Replicas,$
$\quad\quad\ cmd : Commands \cup \{none\},\ deps :$ SUBSET $Instances,\ seq : Nat]$

$SpecReplyMessage\ \triangleq$
$\quad\quad [type : \{\,\text{``spec-reply''}\,\},\ src : Replicas,\ dst : Clients,$
$\quad\quad\ inst : Instances,\ owner : Replicas,$
$\quad\quad\ cmd : Commands \cup \{none\},\ deps :$ SUBSET $Instances,\ seq : Nat,\ committed :$ SUBSET $Instances]$

$CommitFastMessage\ \triangleq$
$\quad\quad [type\ : \{\,\text{``commit-fast''}\,\},\ src : Clients,\ dst : Replicas,$
$\quad\quad\ proof :$ SUBSET $SpecReplyMessage,$
$\quad\quad\ inst : Instances,\ owner : Replicas,$
$\quad\quad\ cmd : Commands \cup \{none\},\ deps :$ SUBSET $Instances,\ seq : Nat]$

$CommitSlowMessage\ \triangleq$
$\quad\quad [type\ : \{\,\text{``commit-slow''}\,\},\ src : Clients,\ dst : Replicas,$
$\quad\quad\ proof :$ SUBSET $SpecReplyMessage,$
$\quad\quad\ inst : Instances,\ owner : Replicas,$
$\quad\quad\ cmd : Commands \cup \{none\},\ deps :$ SUBSET $Instances,\ seq : Nat]$

$CommitReplyMessage\ \triangleq$

$$[\textit{type} : \{\text{``commit-reply''}\}, \textit{src} : \textit{Replicas}, \textit{dst} : \textit{Clients},$$
$$\textit{inst} : \textit{Instances}, \textit{owner} : \textit{Replicas}]$$

$\textit{CommitMessage} \triangleq$
    $\textit{CommitFastMessage} \cup \textit{CommitSlowMessage}$

$\textit{CmdLogEntry} \triangleq [\textit{inst} : \textit{Instances},$
    $\textit{status} : \textit{Status},$
    $\textit{owner} : \textit{Replicas},$
    $\textit{cmd} : \textit{Commands} \cup \{\textit{none}\},$
    $\textit{deps} : \text{SUBSET } \textit{Instances},$
    $\textit{seq} : \textit{Nat},$
    $\textit{proof} : \text{SUBSET } \textit{CommitMessage}]$

$\textit{InitOwnerChgMessage} \triangleq$
    $[\textit{type} \;\; : \{\text{``init-owner-change''}\}, \textit{src} : \textit{Replicas}, \textit{dst} : \textit{Replicas},$
    $\textit{owner} : \textit{Replicas}, \textit{new\_owner} : \textit{Replicas}]$

$\textit{OwnerChgMessage} \triangleq$
    $[\textit{type} \;\; : \{\text{``owner-change''}\}, \textit{src} : \textit{Replicas}, \textit{dst} : \textit{Replicas},$
    $\textit{owner} : \textit{Replicas}, \textit{new\_owner} : \textit{Replicas},$
    $\textit{proof} \;\; : \text{SUBSET } \textit{CmdLogEntry}]$

$\textit{NewOwnerMessage} \triangleq$
    $[\textit{type} \;\; : \{\text{``new-owner''}\}, \textit{src} : \textit{Replicas}, \textit{dst} : \textit{Replicas},$
    $\textit{owner} : \textit{Replicas}, \textit{new\_owner} : \textit{Replicas},$
    $\textit{proof} \;\; : \text{SUBSET } \textit{CmdLogEntry}]$

$\textit{Message} \triangleq$
    $\phantom{\cup\;\;}\textit{SpecOrderMessage}$
    $\cup\;\;\textit{SpecReplyMessage}$
    $\cup\;\;\textit{CommitSlowMessage}$
    $\cup\;\;\textit{CommitFastMessage}$
    $\cup\;\;\textit{CommitReplyMessage}$
    $\cup\;\;\textit{InitOwnerChgMessage}$
    $\cup\;\;\textit{OwnerChgMessage}$
    $\cup\;\;\textit{NewOwnerMessage}$

Variables:

$\textit{cmdLog}\;=\;$ the commands log at each replica
$\textit{proposed}\;=\;$ command that have been proposed
$\textit{executed}\;=\;$ the log of executed commands at each replica
$\textit{sentMsg}\;=\;$ sent (but not yet received) messages
$\textit{crtInst}\;=\;$ the next instance available for a command leader
$\textit{leaderOfInst}\;=\;$ the set of instances each replica has started but not yet finalized

> *committed* = maps commands to set of commit attributs tuples
> *owners* = largest ballot number used by any replica

VARIABLES *cmdLog, proposed, executed, sentMsg, crtInst, leaderOfInst,
committed, ownerLog, owners*

$TypeOK \triangleq$
- $\land \quad cmdLog \in [Replicas \to \text{SUBSET } CmdLogEntry]$
- $\land \quad proposed \in \text{SUBSET } Commands$
- $\land \quad executed \in [Replicas \to \text{SUBSET } (Nat \times Commands)]$
- $\land \quad sentMsg \in \text{SUBSET } Message$
- $\land \quad crtInst \in [Replicas \to Nat]$
- $\land \quad leaderOfInst \in [Replicas \to \text{SUBSET } Instances]$
- $\land \quad committed \in [Instances \to \text{SUBSET } ((Commands \cup \{none\}) \times$
  $(\text{SUBSET } Instances) \times$
  $Nat)]$
- $\land \quad ownerLog \in [Replicas \to Replicas]$
- $\land \quad owners \in [Replicas \to Nat]$

$vars \triangleq \langle cmdLog, proposed, executed, sentMsg, crtInst, leaderOfInst,$
$committed, ownerLog, owners \rangle$

Initial state predicate

$Init \triangleq$
- $\land sentMsg = \{\}$
- $\land cmdLog = [r \in Replicas \mapsto \{\}]$
- $\land proposed = \{\}$
- $\land executed = [r \in Replicas \mapsto \{\}]$
- $\land crtInst = [r \in Replicas \mapsto 1]$
- $\land leaderOfInst = [r \in Replicas \mapsto \{\}]$
- $\land committed = [i \in Instances \mapsto \{\}]$
- $\land ownerLog = [r \in Replicas \mapsto r]$
- $\land owners = [r \in Replicas \mapsto 0]$

Actions

$SpecOrder(cmd, client, cleader, Quorum, inst, oldMsg) \triangleq$
$\quad \text{LET } newDeps \triangleq \{rec.inst : rec \in cmdLog[cleader]\}$
$\quad \quad newSeq \triangleq 1 + Max(\{t.seq : t \in cmdLog[cleader]\})$
$\quad \quad oldRecs \triangleq \{rec \in cmdLog[cleader] : rec.inst = inst\}$ Get any old records
$\quad \quad instCom \triangleq \{t.inst : t \in \{tt \in cmdLog[cleader] :$
$\quad \quad \quad \quad tt.status \in \{\text{"committed"}, \text{"executed"}\}\}\} \text{IN}$
$\quad \land cmdLog' = [cmdLog \text{ EXCEPT } ![cleader] = (@ \setminus oldRecs) \cup$
$\quad \quad \quad \{[inst \mapsto inst,$

4

$$
\begin{aligned}
&status \mapsto \text{``spec-ordered''}, \\
&owner \mapsto cleader, \\
&cmd \mapsto cmd, \\
&deps \mapsto newDeps, \\
&seq \mapsto newSeq, \\
&proof \mapsto \{\}]\}]
\end{aligned}
$$

$\land\ leaderOfInst' = [leaderOfInst \text{ EXCEPT } ![cleader] = @ \cup \{inst\}]$
$\land\ sentMsg' = ((sentMsg \setminus oldMsg) \cup$

$$
\begin{aligned}
[&type\ : \{\text{``spec-order''}\}, \\
&src\ : \{cleader\}, \\
&dst\ : Quorum \setminus \{cleader\}, \\
&inst\ : \{inst\}, \\
&owner : \{cleader\}, \\
&cmd\ : \{cmd\}, \\
&deps\ : \{newDeps\}, \\
&seq\ : \{newSeq\}])
\end{aligned}
$$

$\cup$

$$
\begin{aligned}
\{[&type \mapsto \text{``spec-reply''}, \\
&src \mapsto cleader, \\
&dst \mapsto client, \\
&inst \mapsto inst, \\
&owner \mapsto cleader, \\
&cmd \mapsto cmd, \\
&deps \mapsto newDeps, \\
&seq \mapsto newSeq, \\
&committed \mapsto instCom]\}
\end{aligned}
$$

$StartSpecOrder(C, client, cleader) \triangleq$
  LET $newInst \triangleq \langle cleader, crtInst[cleader] \rangle$
  IN  $\land\ proposed' = proposed \cup \{C\}$
    $\land\ (\exists\ Q \in FastQuorums(cleader) :$
      $SpecOrder(C, client, cleader, Q, newInst, \{\}))$
    $\land\ crtInst' = [crtInst \text{ EXCEPT } ![cleader] = @ + 1]$
    $\land\ \text{UNCHANGED } \langle executed, committed, ownerLog, owners \rangle$

$SpecReply(replica, client, behavior) \triangleq$
  $\land\ \exists\ msg \in sentMsg :$
   $\land\ msg.type = \text{``spec-order''}$
   $\land\ msg.dst = replica$
   $\land\ \text{LET } oldRec \triangleq \{rec \in cmdLog[replica] : rec.inst = msg.inst\} \text{ IN}$
    $\land\ \text{LET } newDeps \triangleq \text{ IF } behavior = \text{``good''} \text{ THEN}$
        $msg.deps\ \cup$
        $(\{t.inst : t \in cmdLog[replica]\} \setminus \{msg.inst\})$
      ELSE $\{\}$
     $newSeq \triangleq \text{ IF } behavior = \text{``good''} \text{ THEN}$

$$\begin{aligned}
&\qquad\qquad Max(\{msg.seq,\\
&\qquad\qquad\qquad 1 + Max(\{t.seq : t \in cmdLog[replica]\})\})\\
&\qquad\qquad \text{ELSE } 1\\
&\qquad instCom \triangleq \{t.inst : t \in \{tt \in cmdLog[replica] :\\
&\qquad\qquad\qquad tt.status \in \{\text{"committed"}, \text{"executed"}\}\}\}\text{IN}\\
&\wedge cmdLog' = [cmdLog \text{ EXCEPT } ![replica] = (@ \setminus oldRec) \cup\\
&\qquad\qquad\qquad \{[inst \mapsto msg.inst,\\
&\qquad\qquad\qquad\ \ status \mapsto \text{"spec-ordered"},\\
&\qquad\qquad\qquad\ \ owner \mapsto msg.owner,\\
&\qquad\qquad\qquad\ \ cmd \mapsto msg.cmd,\\
&\qquad\qquad\qquad\ \ deps \mapsto newDeps,\\
&\qquad\qquad\qquad\ \ seq \mapsto newSeq,\\
&\qquad\qquad\qquad\ \ proof \mapsto \{\}]\}]\\
&\wedge sentMsg' = (sentMsg \setminus \{msg\}) \cup\\
&\qquad\qquad\qquad \{[type \mapsto \text{"spec-reply"},\\
&\qquad\qquad\qquad\ \ src \mapsto replica,\\
&\qquad\qquad\qquad\ \ dst \mapsto client,\\
&\qquad\qquad\qquad\ \ inst \mapsto msg.inst,\\
&\qquad\qquad\qquad\ \ owner \mapsto msg.owner,\\
&\qquad\qquad\qquad\ \ cmd \mapsto msg.cmd,\\
&\qquad\qquad\qquad\ \ deps \mapsto newDeps,\\
&\qquad\qquad\qquad\ \ seq \mapsto newSeq,\\
&\qquad\qquad\qquad\ \ committed \mapsto instCom]\}\\
&\wedge \text{UNCHANGED } \langle proposed, crtInst, executed, leaderOfInst,\\
&\qquad\qquad\qquad committed, ownerLog, owners\rangle
\end{aligned}$$

$$\begin{aligned}
&FastPath(client, cleader, i, Q) \triangleq\\
&\quad \wedge i \in leaderOfInst[cleader]\\
&\quad \wedge Q \in FastQuorums(cleader)\\
&\quad \wedge \text{LET } replies \triangleq \{msg \in sentMsg :\\
&\qquad\qquad\qquad \wedge msg.type = \text{"spec-reply"}\\
&\qquad\qquad\qquad \wedge msg.inst = i\\
&\qquad\qquad\qquad \wedge msg.dst = client\\
&\qquad\qquad\qquad \wedge msg.src \in Q\\
&\qquad\qquad\qquad \wedge msg.owner = cleader\}\text{IN}\\
&\quad \wedge (\forall replica \in Q :\\
&\qquad \exists msg \in replies : msg.src = replica)\\
&\quad \wedge (\forall r1, r2 \in replies :\\
&\qquad \wedge r1.deps = r2.deps\\
&\qquad \wedge r1.seq = r2.seq)\\
&\quad \wedge \text{LET } rep \triangleq \text{CHOOSE } rep \in replies\ \ : \text{TRUE IN}\\
&\qquad \wedge \exists SQ \in SlowQuorums(cleader) :\\
&\qquad\quad (sentMsg' = (sentMsg \setminus replies) \cup\\
&\qquad\qquad [type : \{\text{"commit-fast"}\},\\
&\qquad\qquad\ src : \{client\},
\end{aligned}$$

6

$$
\begin{aligned}
&\phantom{\wedge\ committed' = [}dst : SQ,\\
&\phantom{\wedge\ committed' = [}inst \ \ : \{i\},\\
&\phantom{\wedge\ committed' = [}owner : \{rep.owner\},\\
&\phantom{\wedge\ committed' = [}cmd : \{rep.cmd\},\\
&\phantom{\wedge\ committed' = [}deps : \{rep.deps\},\\
&\phantom{\wedge\ committed' = [}seq : \{rep.seq\},\\
&\phantom{\wedge\ committed' = [}proof : \{replies\}])\\
&\wedge\ committed' = [committed \text{ EXCEPT } ![i] = @\ \cup\\
&\phantom{\wedge\ committed' = [committed }\{\langle rep.cmd,\ rep.deps,\ rep.seq\rangle\}]\\
&\wedge\ \text{UNCHANGED } \langle proposed,\ executed,\ crtInst,\ ownerLog,\ owners,\\
&\phantom{\wedge\ \text{UNCHANGED } \langle}leaderOfInst\rangle
\end{aligned}
$$

$SlowPath(client,\ cleader,\ i,\ Q) \triangleq$
$\quad \wedge\ i\ \in leaderOfInst[cleader]$
$\quad \wedge\ Q \in SlowQuorums(cleader)$
$\quad \wedge\ \text{LET } replies\ \triangleq\ \{msg \in sentMsg :$
$\phantom{\quad \wedge\ \text{LET } replies\ \triangleq\ \{msg}\wedge msg.type = \text{"spec-reply"}$
$\phantom{\quad \wedge\ \text{LET } replies\ \triangleq\ \{msg}\wedge msg.inst = i$
$\phantom{\quad \wedge\ \text{LET } replies\ \triangleq\ \{msg}\wedge msg.dst = cleader$
$\phantom{\quad \wedge\ \text{LET } replies\ \triangleq\ \{msg}\wedge msg.src \in Q$
$\phantom{\quad \wedge\ \text{LET } replies\ \triangleq\ \{msg}\wedge msg.owner = cleader\}\text{IN}$
$\quad\quad \wedge\ (\forall\ replica \in (Q \setminus \{cleader\}):$
$\phantom{\quad\quad \wedge\ (\forall}\exists\ msg \in replies : msg.src = replica)$
$\quad\quad \wedge\ \text{LET } finalDeps\ \triangleq\ \text{UNION }\{msg.deps : msg \in replies\}$
$\phantom{\quad\quad \wedge\ \text{LET }}finalSeq\ \triangleq\ Max(\{msg.seq : msg \in replies\})$
$\phantom{\quad\quad \wedge\ \text{LET }}rep\ \triangleq\ \text{CHOOSE } rep \in replies\ \ : \text{TRUEIN}$
$\quad\quad\quad \wedge\ \exists\ SQ \in SlowQuorums(cleader):$
$\phantom{\quad\quad\quad \wedge\ \exists}(sentMsg' = (sentMsg \setminus replies)\ \cup$
$\phantom{\quad\quad\quad \wedge\ \exists(sentMsg' = }[type : \{\text{"commit-slow"}\},$
$\phantom{\quad\quad\quad \wedge\ \exists(sentMsg' = [}src : \{cleader\},$
$\phantom{\quad\quad\quad \wedge\ \exists(sentMsg' = [}dst : SQ,$
$\phantom{\quad\quad\quad \wedge\ \exists(sentMsg' = [}inst\ \ : \{i\},$
$\phantom{\quad\quad\quad \wedge\ \exists(sentMsg' = [}owner : \{cleader\},$
$\phantom{\quad\quad\quad \wedge\ \exists(sentMsg' = [}cmd : \{rep.cmd\},$
$\phantom{\quad\quad\quad \wedge\ \exists(sentMsg' = [}deps : \{finalDeps\},$
$\phantom{\quad\quad\quad \wedge\ \exists(sentMsg' = [}seq : \{finalSeq\},$
$\phantom{\quad\quad\quad \wedge\ \exists(sentMsg' = [}proof : \{replies\}])$
$\quad\quad \wedge\ \text{UNCHANGED } \langle cmdLog,\ proposed,\ executed,\ crtInst,\ leaderOfInst,$
$\phantom{\quad\quad \wedge\ \text{UNCHANGED } \langle}ownerLog,\ owners,\ committed\rangle$

$FastCommit(replica) \triangleq$
$\quad \exists\ msg \in sentMsg :$
$\quad\quad \wedge\ msg.type = \text{"commit-fast"}$
$\quad\quad \wedge\ msg.dst = replica$
$\quad\quad \wedge\ \text{LET } oldRec\ \triangleq\ \{rec \in cmdLog[replica] : rec.inst = msg.inst\}\text{IN}$

$$
\begin{aligned}
&\wedge\ \mathit{cmdLog}' = [\mathit{cmdLog}\ \text{EXCEPT}\ ![\mathit{replica}] = (@\setminus \mathit{oldRec})\ \cup \\
&\qquad\qquad\{[\mathit{inst}\ \mapsto\ \mathit{msg.inst}, \\
&\qquad\qquad\ \ \mathit{status}\ \mapsto\ \text{``committed''}, \\
&\qquad\qquad\ \ \mathit{owner}\ \mapsto\ \mathit{msg.owner}, \\
&\qquad\qquad\ \ \mathit{cmd}\ \mapsto\ \mathit{msg.cmd}, \\
&\qquad\qquad\ \ \mathit{deps}\ \mapsto\ \mathit{msg.deps}, \\
&\qquad\qquad\ \ \mathit{seq}\ \mapsto\ \mathit{msg.seq}, \\
&\qquad\qquad\ \ \mathit{proof}\ \mapsto\ \{\mathit{msg}\}]\}] \\
&\wedge\ \mathit{sentMsg}' = (\mathit{sentMsg}\setminus\{\mathit{msg}\}) \\
&\wedge\ \text{UNCHANGED}\ \langle \mathit{proposed}, \mathit{crtInst}, \mathit{executed}, \mathit{leaderOfInst}, \\
&\qquad\qquad\qquad \mathit{committed}, \mathit{ownerLog}, \mathit{owners}\rangle
\end{aligned}
$$

$\mathit{SlowCommit}(\mathit{replica})\ \triangleq$
$\quad \exists\,\mathit{msg} \in \mathit{sentMsg} :$
$\qquad \wedge\ \mathit{msg.type}\ =\ \text{``commit-slow''}$
$\qquad \wedge\ \mathit{msg.dst}\ =\ \mathit{replica}$
$\qquad \wedge\ \text{LET}\ \mathit{oldRec}\ \triangleq\ \{\mathit{rec} \in \mathit{cmdLog}[\mathit{replica}] : \mathit{rec.inst} = \mathit{msg.inst}\}\ \text{IN}$
$\qquad\quad \wedge\ (\forall\,\mathit{rec}\ \in \mathit{oldRec} : (\mathit{rec.ballot} = \mathit{msg.ballot} \vee$
$\qquad\qquad\qquad\qquad\qquad \mathit{rec.ballot}[1] < \mathit{msg.ballot}[1]))$
$\qquad\quad \wedge\ \mathit{cmdLog}' = [\mathit{cmdLog}\ \text{EXCEPT}\ ![\mathit{replica}] = (@\setminus \mathit{oldRec})\ \cup$
$\qquad\qquad\qquad \{[\mathit{inst}\ \mapsto\ \mathit{msg.inst},$
$\qquad\qquad\qquad\ \ \mathit{status}\ \mapsto\ \text{``committed''},$
$\qquad\qquad\qquad\ \ \mathit{owner}\ \mapsto\ \mathit{msg.owner},$
$\qquad\qquad\qquad\ \ \mathit{cmd}\ \mapsto\ \mathit{msg.cmd},$
$\qquad\qquad\qquad\ \ \mathit{deps}\ \mapsto\ \mathit{msg.deps},$
$\qquad\qquad\qquad\ \ \mathit{seq}\ \mapsto\ \mathit{msg.seq},$
$\qquad\qquad\qquad\ \ \mathit{proof}\ \mapsto\ \{\mathit{msg}\}]\}]$
$\qquad\quad \wedge\ \mathit{sentMsg}' = (\mathit{sentMsg}\setminus\{\mathit{msg}\})\ \cup$
$\qquad\qquad\qquad \{[\mathit{type}\ \mapsto\ \text{``commit-reply''},$
$\qquad\qquad\qquad\ \ \mathit{src}\ \mapsto\ \mathit{replica},$
$\qquad\qquad\qquad\ \ \mathit{dst}\ \mapsto\ \mathit{msg.src},$
$\qquad\qquad\qquad\ \ \mathit{inst}\ \mapsto\ \mathit{msg.inst},$
$\qquad\qquad\qquad\ \ \mathit{owner}\mapsto\ \mathit{msg.owner}]\}$
$\qquad\quad \wedge\ \text{UNCHANGED}\ \langle \mathit{proposed}, \mathit{crtInst}, \mathit{executed}, \mathit{leaderOfInst},$
$\qquad\qquad\qquad\qquad\qquad \mathit{committed}, \mathit{ownerLog}, \mathit{owners}\rangle$

$\mathit{CommitDone}(\mathit{client}, \mathit{cleader}, i, Q)\ \triangleq$
$\quad \wedge\ i\ \in \mathit{leaderOfInst}[\mathit{cleader}]$
$\quad \wedge\ Q \in \mathit{SlowQuorums}(\mathit{cleader})$
$\quad \wedge\ \text{LET}\ \mathit{replies}\ \triangleq\ \{\mathit{msg} \in \mathit{sentMsg} :$
$\qquad\qquad\qquad\qquad \wedge\ \mathit{msg.type}\ =\ \text{``commit-reply''}$
$\qquad\qquad\qquad\qquad \wedge\ \mathit{msg.inst}\ =\ i$
$\qquad\qquad\qquad\qquad \wedge\ \mathit{msg.dst}\ =\ \mathit{client}$
$\qquad\qquad\qquad\qquad \wedge\ \mathit{msg.src}\ \in\ Q$
$\qquad\qquad\qquad\qquad \wedge\ \mathit{msg.owner}\ =\ \mathit{cleader}\}\ \text{IN}$

∧ (∀ replica ∈ Q : ∃ msg ∈ replies :
            msg.src = replica)
∧ leaderOfInst' = [leaderOfInst EXCEPT ![cleader] = @ \ {i}]
∧ LET rep ≜ CHOOSE rep ∈ replies : TRUE IN
    committed' = [committed EXCEPT ![i] = @ ∪
                        {⟨rep.cmd, rep.deps, rep.seq⟩}]
∧ UNCHANGED ⟨cmdLog, sentMsg, proposed, executed,
              crtInst, ownerLog, owners⟩

Recovery actions

$InitOwnerChange(replica, cleader, nowner, Q) \triangleq$
  ∧ replica ≠ cleader
  ∧ Q ∈ WeakQuorums(nowner)
  ∧ owners[cleader] < (MaxOwners ∗ Cardinality(Q))
  ∧ sentMsg' = sentMsg ∪
              [type       : {"init-owner-change"},
               src        : {replica},
               dst        : Q,
               owner      : {cleader},
               new_owner  : {nowner}]
  ∧ ownerLog' = [ownerLog EXCEPT ![cleader] = nowner]
  ∧ owners' = [owners EXCEPT ![cleader] = @ + 1]
  ∧ UNCHANGED ⟨cmdLog, proposed, executed, crtInst,
              leaderOfInst, committed⟩

$OwnerChange(replica, nowner, Q) \triangleq$
  ∧ Q ∈ WeakQuorums(nowner)
  ∧ replica ∈ Q
  ∧ LET msgs ≜ {msg ∈ sentMsg :
              ∧ msg.type = "init-owner-change"
              ∧ msg.dst = replica
              ∧ msg.new_owner = nowner} IN
    ∧ Cardinality(msgs) ≥ Cardinality(Q)
    ∧ ∀ r1, r2 ∈ msgs :
        ∧ r1.owner = r2.owner
        ∧ r1.new_owner = r2.new_owner
    ∧ LET msg ≜ CHOOSE msg ∈ msgs : TRUE IN
        ∧ (sentMsg' = (sentMsg \ msgs) ∪
                  {[type      ↦ "owner-change",
                    src       ↦ replica,
                    dst       ↦ msg.new_owner,
                    owner     ↦ msg.owner,
                    new_owner ↦ msg.new_owner,

9

$$\qquad\qquad\qquad\qquad proof \;\; \mapsto cmdLog[msg.owner]]\})$$
$\quad\land$ UNCHANGED $\langle cmdLog, proposed, executed, crtInst, leaderOfInst,$
$\qquad\qquad\qquad committed, ownerLog, owners\rangle$

$NewOwner(owner, WQ, SQ) \triangleq$
$\quad\land WQ \in WeakQuorums(owner)$
$\quad\land SQ \;\; \in SlowQuorums(owner)$
$\quad\land$ LET $msgs \triangleq \{msg \in sentMsg :$
$\qquad\qquad\qquad\quad \land msg.type =$ "owner-change"
$\qquad\qquad\qquad\quad \land msg.dst = owner$
$\qquad\qquad\qquad\quad \land msg.new\_owner = owner$
$\qquad\qquad\qquad\quad \land msg.src \in WQ\}$ IN
$\quad\quad \land PrintT("NewOwner - begin")$
$\quad\quad \land PrintT(Cardinality(msgs))$
$\qquad \land Cardinality(msgs) \geq Cardinality(WQ)$
$\quad\quad \land PrintT(Cardinality(msgs))$
$\qquad \land (\forall\, r \in WQ :$
$\qquad\qquad \exists\, msg \in msgs : msg.src = r)$
$\qquad \land \forall\, r1, r2 \in msgs :$
$\qquad\quad \land r1.owner = r2.owner$
$\qquad\quad \land r1.new\_owner = r2.new\_owner$
$\qquad \land \exists\, lmsg \in msgs :$
$\qquad\quad \land \forall\, msg2 \in msgs \setminus \{lmsg\} :$
$\qquad\qquad\quad Cardinality(lmsg.proof) \geq Cardinality(msg2.proof)$
$\qquad\quad \land$ LET $cmsgs \triangleq \{p \in lmsg.proof : p.status =$ "committed"$\}$
$\qquad\qquad\quad\; pmsgs \triangleq \{p \in lmsg.proof : p.status =$ "spec-ordered"
$\qquad\qquad\qquad\qquad \land \;\; Cardinality(p.proof) \geq Cardinality(FakeReplicas) + 1\}$
$\qquad\qquad$ IN
$\qquad\qquad\quad \land (sentMsg' = (sentMsg \setminus msgs) \cup$
$\qquad\qquad\qquad\qquad [type : \{$"new-owner"$\},$
$\qquad\qquad\qquad\qquad\; src : \{owner\},$
$\qquad\qquad\qquad\qquad\; dst : SQ,$
$\qquad\qquad\qquad\qquad\; owner : \{lmsg.owner\},$
$\qquad\qquad\qquad\qquad\; new\_owner : \{lmsg.new\_owner\},$
$\qquad\qquad\qquad\qquad\; proof : cmsgs \cup pmsgs])$      proof: $cmsgs \cup [p$
$\qquad \land$ UNCHANGED $\langle cmdLog, proposed, executed, crtInst,$
$\qquad\qquad\qquad leaderOfInst, committed, ownerLog, owners\rangle$

$CompleteOwnerChange(new, replica, SQ) \triangleq$
$\quad \land SQ \in SlowQuorums(new)$
$\quad \land replica \in SQ$
$\quad \land$ LET $pmsg \triangleq \{msg \in sentMsg :$
$\qquad\qquad\qquad\;\; \land msg.type =$ "new-owner"
$\qquad\qquad\qquad\;\; \land msg.dst = replica$
$\qquad\qquad\qquad\;\; \land msg.src = new\}$ IN

10

$\quad\quad\quad \land\ Cardinality(pmsg) \geq 1$
$\quad\quad\quad \land\ \text{LET}\ msg\ \triangleq\ \text{CHOOSE}\ msg \in pmsg : \text{TRUE} \text{IN}$
$\quad\quad\quad\quad \land\ \text{LET}\ oldRec\ \triangleq\ \{rec \in cmdLog[replica] :$
$\quad\quad\quad\quad\quad\quad \forall\, p \in msg.proof : rec.inst = p.inst\}\text{IN}$
$\quad\quad\quad\quad\quad \land\ cmdLog' = [cmdLog\ \text{EXCEPT}\ ![replica] = msg.proof]$
$\quad\quad\quad\quad\quad \land\ ownerLog' = [ownerLog\ \text{EXCEPT}\ ![msg.owner] = msg.new\_owner]$
$\quad\quad\quad \land\ sentMsg' = sentMsg \setminus \{pmsg\}$
$\quad\quad\quad \land\ PrintT(\text{``CompleteOwnerChange''})$
$\quad\quad\quad \land\ \text{UNCHANGED}\ \langle cmdLog, proposed, executed, crtInst,$
$\quad\quad\quad\quad\quad leaderOfInst, committed, owners\rangle$

### Action groups

$CommandLeaderAction\ \triangleq$
$\quad \lor\ \exists\, cleader \in Replicas :$
$\quad\quad \lor\ \exists\, cmd \in (Commands \setminus proposed) : \exists\, client \in Clients :$
$\quad\quad\quad StartSpecOrder(cmd, client, cleader)$
$\quad \lor\ \exists\, new \in Replicas : \exists\, SQ \in SlowQuorums(new) : \exists\, WQ \in WeakQuorums(new) :$
$\quad\quad NewOwner(new, WQ, SQ)$

$GoodReplicaAction\ \triangleq$
$\quad \lor\ \exists\, replica \in GoodReplicas :$
$\quad\quad \lor\ \exists\, client \in Clients : SpecReply(replica, client, \text{``good''})$
$\quad\quad \lor\ FastCommit(replica)$
$\quad\quad \lor\ SlowCommit(replica)$
$\quad \lor\ \exists\, replica \in Replicas : \exists\, l \in Replicas :$
$\quad\quad \text{LET}\ new\ \triangleq\ NextOwner(l)\text{IN}$
$\quad\quad \lor\ \land\ l \neq new$
$\quad\quad\quad \land\ \exists\, WQ \in WeakQuorums(new) :$
$\quad\quad\quad\quad \lor\ InitOwnerChange(replica, l, new, WQ)$
$\quad\quad\quad\quad \lor\ OwnerChange(replica, new, WQ)$
$\quad\quad \lor\ \exists\, SQ \in WeakQuorums(l) : CompleteOwnerChange(new, replica, SQ)$

$FakeReplicaAction\ \triangleq$
$\quad \lor\ GoodReplicaAction$
$\quad \lor\ \exists\, replica \in FakeReplicas :$
$\quad\quad \exists\, client \in Clients : SpecReply(replica, client, \text{``bad''})$

$ClientAction\ \triangleq$
$\quad \land\ \exists\, client \in Clients : \exists\, replica \in Replicas : \exists\, inst \in leaderOfInst[replica] :$
$\quad\quad \exists\, ballot \in 0\,..\,MaxOwners :$
$\quad\quad\quad \lor\ (\exists\, Q \in FastQuorums(replica) : FastPath(client, replica, inst, Q))$
$\quad\quad\quad \lor\ (\exists\, Q \in SlowQuorums(replica) :$

11

$\qquad\qquad\qquad SlowPath(client, replica, inst, Q))$
$\qquad\qquad \lor (\exists\, Q \in SlowQuorums(replica):$
$\qquad\qquad\qquad CommitDone(client, replica, inst, Q))$
$\quad \land \text{UNCHANGED }\langle cmdLog \rangle$

Next action

$Next \triangleq$
$\quad \lor CommandLeaderAction$
$\quad \lor GoodReplicaAction$
$\quad \lor FakeReplicaAction$
$\quad \lor ClientAction$

The complete definition of the algorithm

$Spec \triangleq Init \land \Box[Next]_{vars}$

Theorems

$Nontriviality \triangleq$
$\quad \forall\, i \in Instances:$
$\qquad \Box(\forall\, C \in committed[i]: C[1] \in proposed \lor C[1] = none)$

$Stability \triangleq$
$\quad \forall\, replica \in Replicas:$
$\qquad \forall\, i \in Instances:$
$\qquad\quad \forall\, C \in Commands:$
$\qquad\qquad \Box((\exists\, rec1 \in cmdLog[replica]:$
$\qquad\qquad\quad \land rec1.inst = i$
$\qquad\qquad\quad \land rec1.cmd = C$
$\qquad\qquad\quad \land rec1.status \in \{\text{``committed''}\}) \Rightarrow$
$\qquad\qquad\quad \Box(\exists\, rec2 \in cmdLog[replica]:$
$\qquad\qquad\qquad \land rec2.inst = i$
$\qquad\qquad\qquad \land rec2.cmd = C$
$\qquad\qquad\qquad \land rec2.status \in \{\text{``committed''}\}))$

$Consistency \triangleq$
$\quad \forall\, i \in Instances:$
$\qquad \Box(Cardinality(committed[i]) \leq 1)$

THEOREM $Spec \Rightarrow (\Box TypeOK) \land Nontriviality \land Stability \land Consistency$

\ ∗ Modification History
\ ∗ Last modified *Tue Apr* 02 13:15:06 *EDT* 2019 by *balaji*
\ ∗ Created *Tue Mar* 19 13:03:49 *EDT* 2019 by *balaji*



\end{document}

%% file: intro.tex
\section{Introduction}

State Machine Replication (SMR) is a common technique employed in today's distributed applications to tolerate server failures and maintain high availability~\cite{Bessani:2014:SMR:2671853.2672428, DBLP:conf/osdi/MaoJM08}. The replication servers, or \emph{replicas}, employ consensus protocols (e.g., Raft~\cite{raft}, Paxos~\cite{lamport2001paxos}, Zab~\cite{zab}) to replicate the application state and ensure that the same sequence of client commands is executed on the shared state in the same order (i.e., a total order), ensuring consistency of the replicated state. 

Consensus solutions can be broadly classified as Crash Fault-Tolerant (CFT) and Byzantine Fault-Tolerant (BFT), with the former being a subset of the latter. While CFT protocols have found their niche in datacenter applications~\cite{spanner, baker2011megastore, cockroachdb} of a single  enterprise-class organization, BFT  protocols are increasingly used in applications involving multiple independent organizations. For example,  for distributed \emph{ledger} or blockchain-based applications in the \emph{permissioned} setting, consensus need to be reached among a set of known participants (e.g., independent organizations), despite no complete trust among them. In such cases, BFT protocols can ensure replica state consistency  while withstanding malicious actors and other non-crash related faults. An  example of a \emph{permissioned} blockchain system that uses a BFT protocol is the Hyperledger Fabric~\cite{androulaki2018hyperledger, Bessani:2017:BFO:3152824.3152830}.

PBFT~\cite{castro1999practical} is, arguably, the first practical implementation of BFT consensus. Since its advent, a number of BFT protocols have been invented. Some reduce client-side latency~\cite{Abd-El-Malek:2005:FBF:1095810.1095817, martin2006fast, kotla2007zyzzyva, cowling2006hq}; some optimize for higher server-side throughput~\cite{amir2008bftattack, Bessani:2014:SMR:2671853.2672428}; 
some have greater tolerance to faults~\cite{li2007beyond}; and some reduce the design and simplementation complexity~\cite{guerraoui2010next}. 


The use of permissioned blockchain applications is growing in many domains (e.g., supply chain~\cite{tian2016supplychain}, real-estate~\cite{spielman2016blockchain}, finance~\cite{eyal2017finance}) and many such systems are increasingly deployed in geographically distributed settings to cater to different application needs.
Geographical-scale (or ``geo-scale'') deployments of BFT systems have an additional challenge: achieving low client-side latencies and high server-side throughput under the high communication latencies of a WAN. Since replicas need to communicate with each other and the clients to reach consensus, the number of communication steps incurred directly impacts the latency, as each step involves sending messages to potentially distant nodes. Thus, protocols such as  
Zyzzyva~\cite{kotla2007zyzzyva}, Q/U~\cite{martin2006fast}, and HQ~\cite{cowling2006hq} use  various techniques to reduce communication steps. These techniques, however, do not reduce client-side latencies in a geo-scale setting, where, the latency per communication step is as important as the number of communication steps. In other words, a protocol can achieve significant cost savings if the latency incurred during a communication step can be reduced. 

The downside of such lack of locality is most manifest for primary-based BFT protocols such as PBFT and Zyzzyva: a replica is bestowed the \emph{primary} status and is responsible for proposing the total-order for all client requests. While the clients that are geographically nearest to the primary may observe optimal client-side latency, the same is not true for distant clients. A distant client will incur higher latency for the first communication step (of sending the request to the primary). Additionally, the primary-based design  limits throughput as the primary carries significantly higher load. 


To validate these hypotheses, we deployed Zyzzyva~\cite{kotla2007zyzzyva} 
in a 4-replica geo-scale setting with nodes located in the US (Virginia), Japan, India, and Australia, using Amazon's EC2 infrastructure~\cite{amazonec2}. 
We deployed clients alongside each replica to inject requests, and measured the client-side latencies by changing the primary's location. Table~\ref{tab:motivation:zyzzyva} shows the results. We observe that the lowest latencies are when the primary is co-located within the same region.

\begin{table}[!ht]
	\centering
	\caption{Zyzzyva's~\cite{kotla2007zyzzyva} latencies (in $ms$) in a geo-scale deployment with primary at different locations. 
	Columns indicate the primary's location. Rows indicate average client-side latency for commands issued from that region. For example, the entry at the $4$th row and the $3$rd  column shows the client-side latency for commands issued from India to the primary in Japan. 
	Lowest latency per primary location is highlighted.
	}
	\label{tab:motivation:zyzzyva}
	\begin{tabular}{|l|c|c|c|c|}
		\hline
		              & Virginia (US) &    Japan     &    India     &  Australia   \\ \hline\hline
		Virginia (US) & \textbf{198}  &     238      &     306      &     303      \\ \hline
		Japan         &      236      & \textbf{167} &     239      &     246      \\ \hline
		India         &      304      &     242      & \textbf{229} &     305      \\ \hline
		Australia     &      303      &     232      &     304      & \textbf{229} \\ \hline
	\end{tabular}
\end{table}

Past work suggests that it is important to perform regular view changes (i.e., elect a new primary) to reduce the negative impacts of a byzantine replica~\cite{veronese2009spin}. This makes it difficult to set latency SLAs for geo-scale applications.

A leaderless protocol can solve the aforementioned problems. A client can send its requests to the nearest replica and can continue to do so as long as the replica is \emph{correct}. The replica can  undertake the task of finding an order among all the concurrent requests in the system, executing the request on the shared state, and return the result. Leaderless protocols~\cite{moraru2013there, caesar, peluso2016making} have been previously proposed for the CFT model. However, to the best of our knowledge, such investigations have not been made for the BFT model. 

%


The transition from CFT to BFT is not straightforward and will yield a sub-optimal solution. As shown in~\cite{lamport2011byzantizing}, additional communication steps as well as more number of messages within each step are fundamental for such transformations. This will result in sub-optimal server-side throughput since more messages should be certified by replicas, (exponentially) increasing the degree of computation. Moreover, additional communication steps will increase client-side latency.

Motivated by these concerns, we present \thesystem, a leaderless BFT  protocol. \thesystem enables every replica in the system to process the requests received from the clients. Doing so  (i) significantly reduces the client-side latency,  (ii) distributes the load across replicas, and (iii) tolerates faults more effectively. Importantly, \thesystem  delivers requests in \emph{three} communication steps in normal operating conditions.

To enable leaderless operation, \thesystem exploits a particular characteristic of client commands: \emph{interference}. In the absence of concurrent interfering commands, \thesystem's clients receive a reply in an optimal three communication steps. When commands interfere, both clients and replicas coherently communicate to establish a consistent total-order, consuming an additional zero or two communication steps.
%
%
\thesystem employs additional techniques such as client-side validation of replica messages and speculative execution to reduce communication steps in the common case, unlike CFT solutions~\cite{moraru2013there, caesar}.

We developed \thesystem's formal specification in TLA+ (available in an accompanying technical report~\cite{techrep}) 
and  show that it provides the classic  properties of BFT protocols including consistency, stability, and liveness. 

To understand how \thesystem fares against state-of-the-art BFT protocols, we implemented \thesystem and conducted an experimental evaluation using the AWS EC2 infrastructure, deploying the  implementations in different sets of geographical  regions. Our evaluation reveals that  \thesystem improves client-side latency by as much as 40\% over  PBFT, FaB, and Zyzzyva.

The paper's central contribution is the \thesystem protocol. To the best of our knowledge, \thesystem is the first BFT protocol to  provide decentralized, deterministic consensus in the eventually synchronous model. By minimizing the latency at each communication step, \thesystem provides a highly effective BFT solution for geo-scale deployments.

The rest of the paper is organized as follows: Section~\ref{sec:model} presents \thesystem's assumptions. Section~\ref{sec:overview} overviews \thesystem, and Section~\ref{sec:design} presents a complete algorithmic design and correctness arguments. An experimental evaluation of \thesystem is presented in Section~\ref{sec:eval}. We summarize past and related work in Section~\ref{sec:relwork} and conclude in Section~\ref{sec:conclusion}.

%% file: model.tex
\section{System Model}
\label{sec:model}

We consider a set of nodes (replicas and clients), in an asynchronous system, that communicate via message passing. The replica nodes have identifiers in the set $\{R_0,...,R_{N-1}\}$. We assume the byzantine fault model in which nodes can behave arbitrarily. We also assume a strong adversary model in which faulty nodes can coordinate to take down the entire system. Every node, however, is capable of producing cryptographic signatures~\cite{Johnson:2001:ECD:2701775.2701951} that faulty nodes cannot break. A message $m$ signed using $R_i$'s private key is denoted as $\langle m \rangle_{R_i}$. The network is fully connected and quasi-reliable: if nodes $R_1$ and $R_2$ are correct, then $p_2$ receives a message from $R_1$ exactly as many times $R_1$ sends it. 

To preserve safety and liveness, \thesystem requires at least $N = 3f + 1$ replica nodes in order to tolerate $f$ Byzantine faults. \thesystem uses two kinds of quorums: a fast quorum with $3f+1$ replicas, and a slow quorum with $2f+1$ replicas. 
Safety is guaranteed as long as only up to $f$ replicas fail. Liveness is guaranteed during periods of synchronous communication.

%% file: overview.tex
\section{Protocol Overview}
\label{sec:overview}

We now overview \thesystem and highlight the novelties that enable it to tolerate byzantine failures and provide optimal wide-area latency.

\thesystem can deliver decisions in three communication steps from the client's point-of-view, if there is no contention, no byzantine failures, and synchronous communication between replicas. The three communication steps include: (i) a client sending a request to any one of the replicas (closest preferably); (ii) a replica forwarding the request to other replicas with a proposed order; and (iii) other replicas (speculatively) executing the request as per the proposed order and replying back to the client. These three steps constitute \thesystem's core novelty. To realize these steps, \thesystem incorporates a set of techniques that we summarize below and explain in detail in the next section.

First, the \thesystem replicas perform speculative execution of the commands after receiving the proposal messages from their respective command-leaders (see below). 
With only one replica-to-replica communication, there is no way to guarantee the final commit order for client commands. Thus, the replica receiving the proposal assumes that other replicas received the same proposal (i.e., the command-leader is not byzantine) and that they have agreed to the proposal. With this assumption, replicas speculatively execute the commands on their local state and return a reply back to the client.

Second, in \thesystem, the client is actively involved in the consensus process. It is responsible for collecting messages from the replicas and ensuring that they have committed to a single order before delivering the reply. The client also enforces a final order, if the replicas are found to deviate.

Third, and most importantly, there are \emph{no} designated primary replicas. Every replica can receive client requests and propose an order for them. To clearly distinguish the replica proposing an order for a command from other replicas, we use the term \emph{command-leader}. A \emph{command-leader} is a replica that proposes an order for the commands received from its clients. For clarity, all replica can be command-leaders. To ensure that client commands are consistently executed across all correct replicas, \thesystem exploits the following concepts.

\thesystem uses the concept of \emph{command interference} 
to empower replicas to make independent commit decisions. If replicas concurrently propose commands that do not interfere, they can be committed and executed independently, in parallel, in any order, and without the knowledge of other non-interfering commands. However, when concurrent commands do interfere, replicas must settle on a common sequential execution. 

\subsubsection*{Command Interference} 
A command encapsulates an operation that must be executed on the shared state. We say that two commands $L_0$ and $L_1$ are interfering if the execution of these commands in different orders on a given state will result in two final states. That is, if there exists a sequence of commands $\Sigma$ such that the serial execution of $\Sigma,L_0,L_1$ is not equivalent to $\Sigma,L_1,L_0$, then $L_0$ and $L_1$ are interfering. 

\subsubsection*{Instance Space} An instance space can be visualized as a sequence of numbered slots to which client-commands can be associated with. The sequence defines the execution order of requests, and the role of a consensus protocol is to reach agreement among a set of replicas on a common order. However, to accommodate our requirement that every replica can be a command-leader for their received requests, each replica has its own instance space. Thus,  \thesystem's role is not only to reach consensus on the mapping of client-commands to the slots within an instance space, but also among the slots in different instance spaces. 

\subsubsection*{Instance Number} An instance number, denoted $I$, is a tuple of the instance space (or replica) identifier and a slot identifier. 

\subsubsection*{Instance Owners} An owner number, $O$, is a monotonically increasing number that is used to identify the owner of an instance space. Thus, there are as many owner numbers as there are instance spaces (or replicas). This number becomes useful when a replica is faulty, and its commands must be recovered by other replicas. When replicas fail, another correct replica steps up to take ownership of the faulty replica's instance space. The owner of a replica $R_0$'s instance space can be identified from its owner number using the formula $O_{R0}$ mod $N$, where $N$ is the number of replicas.
Initially, the owner number of each instance space is set to the owner replica's identifier (e.g., $O_{R0} = 0$, $O_{R1} = 1$, and so on).

\subsubsection*{Dependencies} Due to the use of per-replica instance spaces, the protocol must agree on the relationship between the slots of different instances spaces. \thesystem does this via dependency collection, which uses the command interference relation. The dependency set $\mathcal{D}$ for command $L$ is every other command $L'$ that interferes with $L$. 

\subsubsection*{Sequence Number ($\mathcal{S}$)} is a globally shared number that is used to break cycles in dependencies. It starts from one and is always set to be larger than all the sequence numbers of the interfering commands. Due to concurrency, it is possible that interfering commands originating from different command-leaders are assigned the same sequence number. In such cases, the replica identifiers are used to break ties.

\subsubsection*{Protocol Properties} 

\thesystem has the following properties:
\begin{enumerate}
	\item \textbf{Nontriviality.} Any request committed and executed by a replica must have been proposed by a client.
	\item \textbf{Stability.} For any replica, the set of committed requests at any time is a subset of the committed requests at any later time. If at time $t_1$, a replica $R_i$ has a request $L$ committed at some instance $I_L$, then $R_i$ will have $L$ committed in $I_L$ at any later time $t_2 > t_1$. 
	\item \textbf{Consistency.} Two replicas can never have different requests committed for the same instance.
	\item \textbf{Liveness.} Requests will eventually be committed by every non-faulty replica, as long as at least $2f+1$ replicas are correct.
\end{enumerate}

%% file: design.tex
\section{Protocol Design}
\label{sec:design}

In this section, we present \thesystem in detail, along with an informal proof of its properties. We have also developed a TLA+ specification of \thesystem and model-checked the protocol correctness; this can be found in the technical report~\cite{techrep}. 

A client command may either take a \emph{fast path} or a \emph{slow path}. The fast path consists of three communication steps, and is taken under no contention, no byzantine failures, and during synchronous communication periods. The \emph{slow path} is taken otherwise to guarantee the final commit order for commands, and incurs two additional steps. 

\subsection{The Fast Path Protocol}

The fast path consists of three communication steps in the critical path and one communication step in the non-critical path of the protocol. Only the communication steps in the critical path contribute to the client-side latency. The fast path works as follows.

\begin{mdframed}[style=box,nobreak=true,align=center]
1. Client sends a request to a replica.
\end{mdframed}

The client $c$ requests a command $L$ to be executed on the replicated state by sending a message $\langle \textsc{Request}, L, t, c\rangle_{\sigma_c}$ to a \thesystem replica. The closest replica may be chosen to achieve the optimal latency. The client includes a timestamp $t$ to ensure exactly-once execution.

\begin{mdframed}[style=box,nobreak=true,align=center]
2. Replica receives a request, assigns an instance number, collects dependencies and assigns a sequence number, and forwards the request to other replicas. 
\end{mdframed}

When a replica $R_i$ receives the message $m = \langle \textsc{Request}, L,$ $t, c\rangle_{\sigma_c}$, it becomes the command-leader for $L$. It assigns $c$ to the lowest available instance number $I_L$ in its instance space and collects a dependency set $\mathcal{D}$ using the command interference relation that was previously described. A sequence number $\mathcal{S}$  assigned for $c$ is calculated as the maximum of sequence numbers of all commands in the dependency set. This information is relayed to all other replicas in a message $\langle\langle \textsc{SpecOrder},$ $O_{Ri}, I_L, \mathcal{D}, \mathcal{S}, h, d \rangle_{\sigma_{Ri}}, m\rangle$, where $d = H(m)$, $h$ is the digest of $R_i$'s instance space, and $O_{Ri}$ is its owner number.

\underline{\emph{Nitpick}}. Before taking the above actions, $R_i$ ensures that the timestamp $t > t_c$, where $t_c$ is the highest time-stamped request seen by  $R_i$ thus far. If not, the message is dropped.

\begin{mdframed}[style=box,nobreak=true,align=center]
3. Other replicas receive the \textsc{SpecOrder} message, speculatively executes the command according to its local snapshot of dependencies and sequence number, and replies back to the client with the result and an updated set of dependencies and sequence number, as necessary.
\end{mdframed}

When replica $R_j$ receives a message $\langle\langle \textsc{SpecOrder}, O_{Ri},$ $I_L, \mathcal{D}, \mathcal{S}, h, d \rangle_{\sigma_{Ri}}, m\rangle$ from replica $R_i$, it ensures that $m$ is a valid \textsc{Request} message and that $I_L = maxI_{Ri} + 1$, where $maxI_{Ri}$ is the largest occupied slot number in $R_i$'s instance space. Upon successful validation, $R_j$ updates the dependencies and sequence number according to its log, speculatively executes the command, and replies back to the client. A reply to the client consists of a message $\langle\langle \textsc{SpecReply},$ $O_{Ri}, I_L, \mathcal{D}', \mathcal{S}', d, c, t \rangle_{\sigma_{Ri}}, R_j, rep, SO\rangle$, where $rep$ is the result, and $SO=\langle \textsc{SpecOrder}, O_{Ri}, I_L,$ $\mathcal{D}, \mathcal{S}, h, d \rangle_{\sigma_{Ri}}$.

\begin{mdframed}[style=box,nobreak=true,align=center]
4. The client receives the speculative replies and dependency metadata.
\end{mdframed}

The client receives messages $\langle\langle \textsc{SpecReply}, O, I_L, \mathcal{D}', \mathcal{S}', d, c, t \rangle_{\sigma_{Rk}}, R_k, rep, SO\rangle$, where $R_k$ is the sending replica. The messages from different replicas are said to match if they have identical $O$, $I_L$, $\mathcal{D}'$, $\mathcal{S}'$, $c$, $t$, and $rep$ fields. The number of matched responses decides the fast path or slow path decision for command $L$.

\begin{mdframed}[style=box,nobreak=true,align=center]
4.1 The client receives $3f+1$ matching responses.
\end{mdframed}

The receipt of $3f+1$ matching responses from the replicas constitutes a fast path decision for command $L$. This happens in the absence of faults, network partitions, and contention. The client returns reply $rep$ to the application and then asynchronously sends a message $\langle \textsc{CommitFast}, c, I_L, CC \rangle$, where $CC$ is the commit certificate consisting of $3f+1$ matching \textsc{SpecReply} responses, and returns.

\begin{mdframed}[style=box,nobreak=true,align=center]
5. The replicas receive either a \textsc{CommitFast} or a \textsc{Commit} message.
\end{mdframed}

\begin{mdframed}[style=box,nobreak=true,align=center]
5.1 The replicas receive a \textsc{CommitFast} message.
\end{mdframed}

Upon receipt of a $\langle \textsc{CommitFast}, c, I_L, CC \rangle$ message, the replica $R_i$ marks the state of $L$ as committed in its local log and enqueues the command for final execution. The replica does not reply back to the client.

\begin{figure}
	\includegraphics[width=0.48\textwidth]{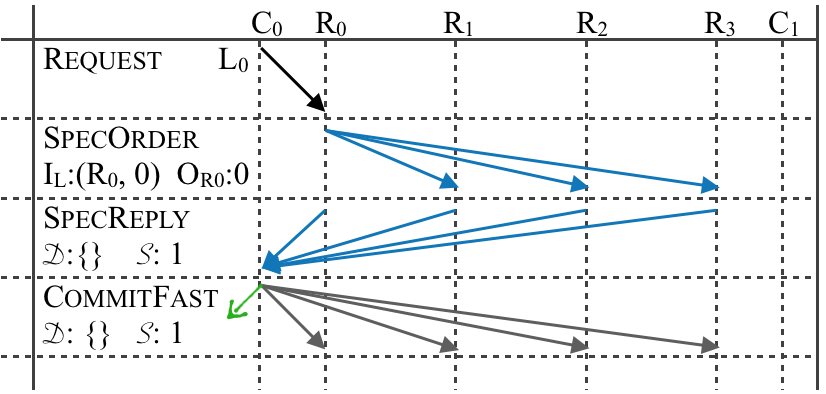}
	\caption{An example of a fast path execution.}
	\label{fig:example-fp}
\end{figure}

\textbf{Example}. Figure~\ref{fig:example-fp} shows an example case. The client sends a signed \textsc{Request} message to replica $R_0$ to execute a command $L_0$ on the replicated service. Replica $R_0$ assigns the lowest available instance number in its instance space to $L_0$. Assuming that no instance number was used previously, the instance number assigned to $L_0$ is $I_{L_0} = \langle r_0, 0 \rangle$. Then, $R_0$ collects dependencies and assigns a sequence number to $L_0$. As the first command, there exists no dependencies, so the dependency set $\mathcal{D} = \{\}$. Thus, the sequence number is $\mathcal{S} = 1$. 

A signed \textsc{SpecOrder} message is sent to other replicas in the system with the command and compiled metadata. Other replicas -- $R_1$ through $R_3$ -- receive this message, add the command to their log, and start amassing dependencies from their log that are not present in $\mathcal{D}$. No other replica received any other command, thus they produce an empty dependency set as well, and  the sequence number remains the same. Since there are no dependencies, all replicas immediately execute the command,  speculatively, on their copy of the application state. The result of execution, unchanged dependency set, sequence number, and the digest of log are sent in a \textsc{SpecReply} message to the client. The client checks for identical replies and returns the result to the application. The replies are identical in this case because no other command conflicts with $L_0$ in any of the replicas and the replicas are benign.

\subsection{Execution Protocol}

\thesystem uses \emph{speculative execution} as a means to reply to the client quickly. However, the protocol must ensure that every correct replica has identical copies of the state. This means that, when necessary (as described in Section~\ref{sec:design:slowpath}), the speculative state must be rolled back and the command must be re-executed correctly; this is called \emph{final execution}.

Moreover, differently from existing BFT solutions, \thesystem collects command dependencies that form a directed graph with potential cycles. The graph must be processed to remove cycles and 
obtain the 
execution order for a command. 

Each replica takes the following steps:
\begin{compactenum}
	\item Waits for the command to be enqueued for execution. For final execution, wait for the dependencies to be committed and enqueued for final execution as well.
	\item A dependency graph is constructed by including $R$ and all its dependencies in $\mathcal{D}$ as nodes and adding edges between nodes indicating the dependencies. The procedure is repeated recursively for each dependency.
	\item Strongly connected components are identified and sorted topologically. 
	\item Starting from the inverse topological order, every strongly connected component is identified, and all the requests within the component are sorted in the sequence number order. The requests are then executed in the sequence number order, breaking ties using replica identifiers.  During speculative execution, the execution is marked \emph{speculative} on the shared state. 
	During final execution, the speculative results are invalidated, command re-executed, and marked \emph{final}. 
\end{compactenum}

Note that speculative execution can happen in either the  \emph{speculative} state or in the \emph{final} version of the state, which ever is the latest. However, for final execution, commands are  executed only on the previous \emph{final} version.

\subsection{The Slow Path Protocol}
\label{sec:design:slowpath}

The slow path is triggered whenever a client receives either unequal and/or insufficient \textsc{SpecReply} messages that is necessary to guarantee a fast path. The client will receive unequal replies if the replicas have different perspectives of the command dependencies, possibly due to contention or due to the presence of byzantine replicas. The case of insufficient replies happen due to  network partitions or byzantine replicas. 

The steps to commit a command in the slow path are as follows.

\begin{mdframed}[style=box,nobreak=true,align=center]
	4.2 The client receives at least $2f+1$ possibly unequal responses.
\end{mdframed}

The client $c$ sets a timer as soon as a \textsc{Request} is issued. When the timer expires, if $c$ has received at least $2f+1$ $\langle\langle \textsc{SpecReply}, O_{Ri}, I_L, \mathcal{D}, \mathcal{S}, d, c, t \rangle_{\sigma_{Ri}}, R_j, rep, SO\rangle$ messages, it produces the final dependency set and sequence number for $L$. The dependency sets from a known set of $2f+1$ replicas are combined to form $\mathcal{D}'$. A new sequence number $\mathcal{S}'$ is generated if the individual dependency sets were not equal. $c$ sends a $\langle \textsc{Commit}, c, I_L, \mathcal{D}', \mathcal{S}', CC \rangle_{\sigma_c}$ message to all the replicas, where $CC$ is the commit certificate containing $2f+1$ \textsc{SpecReply} messages that are used to produce the final dependency set.

\underline{\emph{Nitpick}}. Each command-leader specifies a known set of $2f+1$ replicas that will form the slow path quorum, which is used by the client to combine dependencies when more than $2f+1$ reply messages are received. This information is relayed to the clients by the respective command-leaders and is cached at the clients.

\begin{mdframed}[style=box,nobreak=true,align=center]
	5.2 The replicas receive a \textsc{Commit} message.
\end{mdframed}

Upon receipt of a $\langle \textsc{Commit}, c, I_L, \mathcal{D}', \mathcal{S}', CC \rangle_{\sigma_c}$  message, replica $r$ updates command $L$'s metadata with the received dependency set $\mathcal{D}'$ and sequence number $\mathcal{S}'$. The state produced after the speculative execution of $L$ is invalidated, and $L$ is enqueued for final execution. 
The result of final execution, $rep$ is sent back to the client in a $\langle \textsc{CommitReply}, L, rep \rangle$ message.

\begin{mdframed}[style=box,nobreak=true,align=center]
	6.2 The client receives $2f+1$ \textsc{CommitReply} messages and returns to the application.
\end{mdframed}

The client returns $rep$ to the application upon receipt of $2f+1$ $\langle \textsc{CommitReply}, L, rep \rangle$ messages. At this point, execution of command $L$ is guaranteed to be safe in the system, while tolerating upto $f$ byzantine failures. Moreover,  even after recovering from failures, all correct replicas will always execute $L$ at this same point in their history to produce the same result.

\begin{figure}
	\centering
	\includegraphics[width=0.45\textwidth]{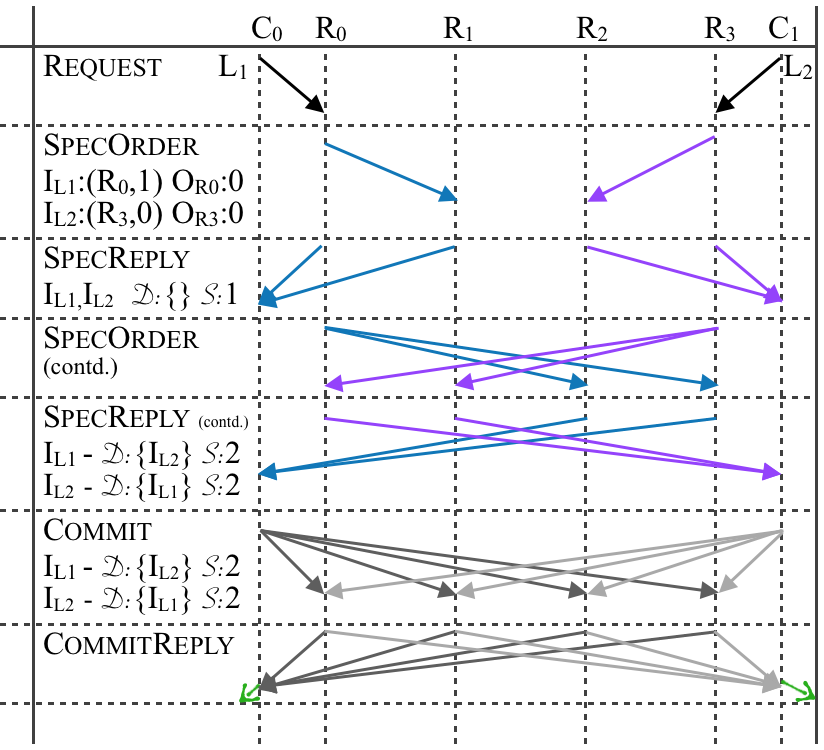}
	\caption{\thesystem: An example of a slow path execution.}
	\label{fig:example-sp}
\end{figure}

\subsubsection*{Example}  Figure~\ref{fig:example-sp} shows an example of a slow path. Two clients $c_0$ and $c_1$ send signed \textsc{Request} messages to replicas $R_0$ and $R_3$, respectively, to execute commands $L_1$ and $L_2$, respectively, on the replicated service. Assume that $L_1$ and $L_2$ conflict. Replica $R_0$ assigns the lowest available instance number of $\langle R_0, 0 \rangle$ to $L_1$. Thus, $R_0$ collects a dependency set $\mathcal{D}_{L_1} = \{\}$ and assigns a sequence number $ \mathcal{S}_{L_1} = 1$ to $L_1$. Meanwhile, $R_3$ assigns the instance number $\langle R_3, 0 \rangle$ to $L_1$; 
the dependency set is $\mathcal{D}_{L_2} = \{\}$, and sequence number is $\mathcal{S}_{L_2} = 1$. Replicas $R_0$ and $R_3$ send \textsc{SpecOrder} messages with their respective commands and their metadata to other replicas. Let's assume that $R_0$ and $R_1$ receive $L_1$ before $L_2$, while $R_2$ and $R_3$ receive $L_2$ before $L_1$. The dependency set and sequence number will remain unchanged for $L_1$ at $R_0$ and $R_1$, because the dependency set in the \textsc{SpecOrder} message received is the latest. However, the dependency set and sequence number for $L_2$ will update to $\mathcal{D}_{L_2}' = \{L_1\}$ and $\mathcal{S}_{L_2}' = 2$, respectively. Similarly, the dependency set and sequence number will remain unchanged for $L_2$ at $R_0$ and $R_1$, but for $L_1$, $\mathcal{D}_{L_1}' = \{L_2\}$ and $\mathcal{S}_{L_1}' = 2$, respectively. The \textsc{SpecReply} messages for both $L_1$ and $L_2$ with the new metadata are sent to the respective clients $c_0$ and $c_1$ by the replicas. 

Let's assume that the slow quorum replicas are $R_0$, $R_1$, and $R_2$ for $R_0$, and $R_1$, $R_2$, and $R_3$ for $R_3$. Since client $c_0$ observes unequal responses, it combines the dependencies for $L_1$ from the slow quorum and selects the highest sequence number to produce the final dependency set $\mathcal{D}_{L_1} = \{L_2\}$ and sequence number $\mathcal{S}_{L_1} = 2$. This metadata is sent to the replicas in a \textsc{Commit} message. Similarly, $c_1$ produces the final dependency set $\mathcal{D}_{L_2} = \{L_1\}$ and sequence number $\mathcal{S}_{L_2} = 2$ for $L_2$, and sends a \textsc{Commit} message to the replicas. The replicas update the dependency set and sequence number of the commands upon receipt of the respective \textsc{Commit} messages, and the commands are queued for execution. The replicas wait for the receipt of the  \textsc{Commit} messages of all commands in the dependency set before processing them. 

After the construction of the graph and the inverse topological sorting, there will exist commands $L_1$ and $L_2$ with a cyclic dependency between them. Since the sequence numbers for both the commands are the same and thus cannot break the dependency, the replica IDs are used in this case. Thus, $L_1$ gets precedence over $L_2$. $L_1$ is executed first, followed by $L_2$. The result of the executions are sent back to the clients. The clients collect $2f+1$ reply messages and return the result to the application.

\begin{figure}
	\centering
	\includegraphics[width=0.45\textwidth]{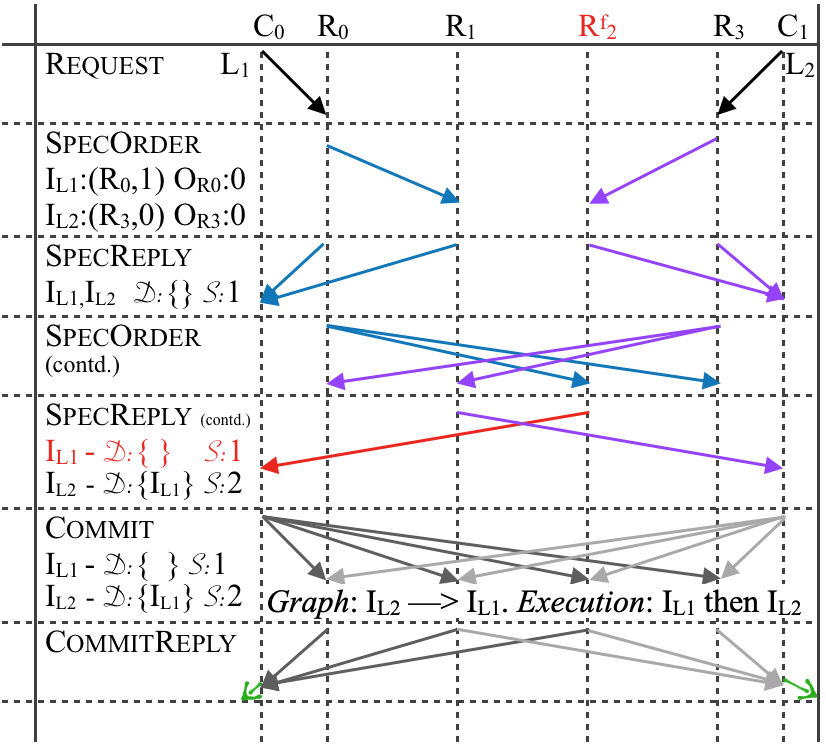}
	\caption{An example of a slow path with a faulty replica.}
	\label{fig:example-spfaulty}
\end{figure}

\subsubsection*{Example with a faulty replica} Figure~\ref{fig:example-spfaulty} shows an example of the slow path that is very similar to that of Figure~\ref{fig:example-sp}, 
but with a faulty replica $R_2$ that misbehaves. Notice that the \textsc{Request} and \textsc{SpecOrder} steps (first four rows) remain the same. Upon receipt of the \textsc{SpecOrder} message from $R_0$ and $R_3$ for $L_1$ and $L_2$, respectively, the replicas collect the dependency set and update the sequence number, and send back \textsc{SpecReply} messages to the client. For $L_1$, $R_0$ and $R_1$ send $\mathcal{D}_{L_1}' = \{\}$ and $\mathcal{S}_{L_1}' = 1$; however, $R_2$ misbehaves and sends $\mathcal{D}_{L_1}' = \{\}$ and $\mathcal{S}_{L_1}' = 1$, even though it received $L_2$ before $L_1$. For $L_2$, $R_2$ and $R_3$ send $\mathcal{D}_{L_2}' = \{\}$ and $\mathcal{S}_{L_2}' = 1$; $R_1$ sends $\mathcal{D}_{L_2}' = \{L_1\}$ and $\mathcal{S}_{L_2}' = 2$. $c_0$ receives a quorum of $2f+1$ \textsc{SpecReply} messages, and sends a \textsc{Commit} message with an empty dependency set and a sequence number 1. On the other hand, $R_1$ that participated in command $L_1$'s quorum sends back a correct dependency set and sequence number. Therefore, the final commit message for $L_2$ will have $L_1$ in its dependency set. Thus, even though replicas immediately execute $L_1$ since $L_1$'s final dependency set is empty, 
they cannot do so for $L_2$. Correct replicas must wait until $L_1$ is \emph{committed} before constructing the graph, at which point $L_1$ will be executed first because of the smallest sequence number, followed by $L_2$.

\subsection{Triggering Owner Changes}

\thesystem employs a mechanism at the clients to monitor the replicas and take actions to restore the service when progress is not being made. Although the slow path helps overcome the effects of a participant byzantine replica, it does ensure progress of a command when its command-leader, the replica that proposed that command, is byzantine. From the client-side, two events can be observed to identify misbehaving command-leaders.

\begin{mdframed}[style=box,nobreak=true,align=center]
	4.3 The client times-out waiting for reply from the replicas.
\end{mdframed}

After the client sends a request with command $L$, it starts another timer, in addition to the one for slow-path, waiting for responses. If the client receives zero or fewer than $2f+1$ responses within the timeout, it sends the $\langle \textsc{Request}, L, t, c, R_i\rangle_{\sigma_c}$ message to all the replicas, where $R_i$ is the original recipient of the message.

When replica $R_j$ receives the message, it takes one of the following two actions. 
If the request matches or has a lower timestamp $t$ than the currently cached timestamp for $c$, then the cached response is returned to $c$. Otherwise, the replica sends a $\langle \textsc{ResendReq}, m, R_j \rangle$ message where $m = \langle \textsc{Request}, L, t, c, R_i\rangle_{\sigma_c}$ to $R_i$ and starts a timer. If the timer expires before the receipt of a \textsc{SpecOrder} message, $R_j$ initiates an ownership change.

\begin{mdframed}[style=box,nobreak=true,align=center]
	4.4 The client receives responses indicating inconsistent ordering by the command-leader and sends a proof of misbehavior to the replicas to initiate an ownership change for the command-leader's instance space.
\end{mdframed}

Even though a client may receive prompt replies from the replicas, it must check for inconsistencies leading to a \emph{proof of misbehavior} against the command-leader. The  $\langle\langle \textsc{SpecReply}, O_{Ri}, I_L, \mathcal{D}', \mathcal{S}', d, c, t \rangle_{\sigma_{Ri}}, R_k, rep, SO\rangle$ messages from different replicas are said to match if they have identical $O_{Ri}$, $I$, $\mathcal{D}$, $\mathcal{S}$, $c$, $t$, and $rep$ fields, but the contention may affect the equality of the dependency set and sequence number fields. Thus, the command-leader is said to misbehave if it sends \textsc{SpecOrder} messages with different instance numbers to different replicas (i.e., if the $I$ field varies between the replicas). The client $c$ can identify by inspecting \textsc{SpecOrder} $SO$ message embedded in the \textsc{SpecReply} message received from the replicas.
In this case, the client collects a pair of such messages to construct a $\langle \textsc{POM}, O_{Ri}, POM \rangle$ message, where $POM$ is a pair of \textsc{SpecReply} messages, proving misbehavior by the command-leader $R_i$ of $L$.

\subsection{The Owner Change Protocol}



An ownership change is triggered for an instance space if its original owner is \emph{faulty}. However, to initiate an ownership change, there must exist either a proof of misbehavior 
against the owner, or enough replicas must have timed out waiting for a reply from the owner.

A replica $R_j$ commits to an ownership change by sending a $\langle \textsc{StartOwnerChange}, R_i, O_{Ri} \rangle$ message to other replicas, where $R_i$ is the suspected replica and $O_{Ri}$ is its owner number. 

When another replica $R_k$ receives at least $f+1$ \textsc{StartOwnerChange} messages for $R_i$, it commits to an ownership change. From this point forward, $R_k$ will not participate in $R_i$'s instance space. The new owner number is calculated as $O_{Ri}' = O_{Ri} + 1$, and the new command-leader is identified using $O_{Ri}'$ $mod$ $N$ (henceforth $R_{l}$). Replicas that have committed to an owner-change send $\langle \textsc{OwnerChange} \rangle$ messages to the new leader. Once the new command-leader $R_l$ receives $f+1$ \textsc{OwnerChange} messages, it becomes the new owner of $R_i$'s instance space and finalizes its history.

Each replica sends an \textsc{OwnerChange} message containing its view of $R_i$'s instance space, i.e., the instances (speculative) executed or committed since the last checkpoint, and the commit-certificate with the highest owner number that it had previously responded to with a commit message, if any. The new owner collects a set $P$ of \textsc{OwnerChange} messages 
and selects only the one that satisfies one of the following conditions. For clarity, we label the sequence of instances in each \textsc{OwnerChange} message as $P_i$, $P_j$, and so on.

There exists a sequence $P_i$ that is the longest and satisfies one of the following conditions.
\begin{description}
	\item[\namedlabel{list:correctness:1}{Condition 1}] $P_i$ has \textsc{Commit} messages with the highest owner number to prove its entries.
	\item[\namedlabel{list:correctness:2}{Condition 2}] $P_i$ has at least $f+1$ \textsc{SpecOrder} messages with the highest owner number to prove its entries.
\end{description}

If there exists a sequence $P_j$ that extends a $P_i$ satisfying any of the above conditions, then $P_j$ is a valid extension of $P_i$ if one of the following conditions hold:
\begin{compactenum}
	\item $P_i$ satisfies~\ref{list:correctness:1}, and for every command $L$ in $P_j$ not in $P_i$, $L$ has at least $f+1$ \textsc{SpecOrder} messages with the same highest order number as $P_i$.
	\item $P_i$ satisfies~\ref{list:correctness:2}, and for every command $L$ in $P_j$ not in $P_i$, $L$ has a signed \textsc{Commit} message with the same highest order number as $P_i$.
\end{compactenum}


The new owner sends a \textsc{NewOwner} message to all the replicas. The message includes the new owner number $O_{R0}'$, the set $P$ of \textsc{OwnerChange} messages that the owner collected as a proof, and the set of safe instances $G$ produced using~\ref{list:correctness:1} and~\ref{list:correctness:2}. 
A replica accepts a \textsc{NewOwner} message if it is valid, and applies the instances from $G$ in $R_i$'s instance space. If necessary, it rolls-back the  speculatively executed requests and re-executes them again.

At this point, $R_i$'s instance space is frozen. No new commands are ordered in the instance space, because each replica has its own instance space that it can use to propose its command. The owner change is used to ensure the safety of commands proposed by the faulty replicas.



%% file: correctness.tex
\subsection{Correctness}
\label{sec:proof}

We formally specified \thesystem in TLA+ and model-checked using  the TLC model checker. The TLA+ specification is provided in a  technical report~\cite{techrep}. In this section, we provide an intuition of how \thesystem achieves its properties. 

\textbf{Nontriviality.} Since clients must sign the requests they send, a malicious primary replica cannot modify them without being suspected. Thus, replicas only execute requests proposed by clients.

\textbf{Consistency.} To prove consistency, we need to show that if a replica $R_j$ commits $L$ at instance $I$, then for any replica $R_k$ that commits $L'$ at $I$, $L$ and $L'$ must be the same command.

To prove this, consider the following. If $R_j$ commits $L$ at $I = \langle R_i, - \rangle$, then an order change should have been executed for replica $R_i$'s instance space. If $R_j$ is correct, then it would have determined that $L$ was executed at $I$ using the commit certificate in the form of \textsc{SpecOrder} or \textsc{Commit} messages embedded within the \textsc{ChangeOwner} messages.
Thus, $L$ and $L'$ must be the same. If $R_j$ is malicious, then the correct replicas will detect it using the invalid progress-certificate received. This will cause an ownership change.

In addition, we need to also show that conflicting requests $L$ and $L'$ are committed and executed 
in the same order across all correct replicas. Assume that $L$ commits with $\mathcal{D}$ and $\mathcal{S}$, while $L'$ commits with $\mathcal{D}'$ and $\mathcal{S}'$. If $L$ and $L'$ conflict, then at least one correct replica must have responded to each other in the dependency set among a quorum of $2f+1$  replies received by the client. Thus, $L$ will be in $L'$'s dependency set and/or viceversa. The execution algorithm is deterministic, and it uses the sequence number to break ties. Thus, conflicting requests will be executed in the same order across all correct replicas.

\textbf{Stability.} Since only $f$ replicas can be byzantine, there must exist at least $2f +1$ replicas with the correct history. During an ownership change, $2f + 1$ replicas should send their history to the new owner which then validates it. Thus, if a request is committed at some instance, it will be extracted from history after any subsequent owner changes and 
committed at same instance at all correct replicas. 

\textbf{Liveness.} Liveness is guaranteed as long as fewer than $f$ replicas crash. Each primary replica attempts to take the fast path with a quorum of $3f + 1$ replicas. When faults occur and a quorum of $3f + 1$ replicas is not possible, the client pursues the slow path with $2f+1$ replicas and  terminates in two additional communication steps.


%% file: evaluation.tex
\section{Evaluation}
\label{sec:eval}

We implemented \thesystem, and its state-of-the-art competitors PBFT, FaB, and Zyzzyva in Go, version 1.10. 
In order to evaluate all systems in a common framework, we used \texttt{gRPC}~\cite{grpc} for communication and \texttt{protobuf}~\cite{protobuf} for message serialization. 
We used the HMAC~\cite{krawczyk1997hmac} and ECDSA~\cite{Johnson:2001:ECD:2701775.2701951} algorithms in Go's \emph{crypto} package to authenticate the messages exchanged by the clients and the replicas. The systems were deployed in different Amazon Web Service (AWS) regions using the EC2 infrastructure~\cite{amazonec2}. 
The VM instance used was m4.2xlarge with 8 vCPUs and 32GB of memory, running Ubuntu 16.04. We implemented a replicated key-value store to evaluate the protocols. Note that, for Zyzzyva and \thesystem, the client process implements the logic for the client portion of the respective protocols.

Among the protocols evaluated, only \thesystem is affected by contention. Contention, in the context of a replicated key-value store, is defined as the percentage of requests that concurrently access the same key. Prior work~\cite{moraru2013there} has shown that, in practice, contention is usually between 0\% and 2\%. Thus, a 2\% contention means that roughly 2\% of the requests issued by  clients target the same key, and the remaining requests target clients' own (non-overlapping) set of keys. However, we evaluate \thesystem at higher contention levels for completeness.

%

\subsection{Client-side Latency}

To understand \thesystem's effectiveness in achieving optimal latency at each geographical region, we devised two experiments to measure the average latency experienced by clients located at each region for each of the protocols. 

\subsubsection*{Experiment 1}
\label{sec:eval:exp1}
We deployed the protocols with four replica nodes in the AWS regions: US-East-1 (Virginia), India, Australia, and Japan.
At each node, we also co-located a client that sends requests to the replicas. For single primary-based protocols (PBFT, FaB, Zyzzyva), 
the primary was set to US-East replica; thus, clients in other replicas send their requests to the primary. For \thesystem, the client sends its requests to the nearest replica (which is in the same region). The clients send requests in closed-loop, meaning that a client will wait for a reply to its previous request before sending another one.

\begin{figure}[htbp]
	\centering
	\includegraphics[width=\linewidth]{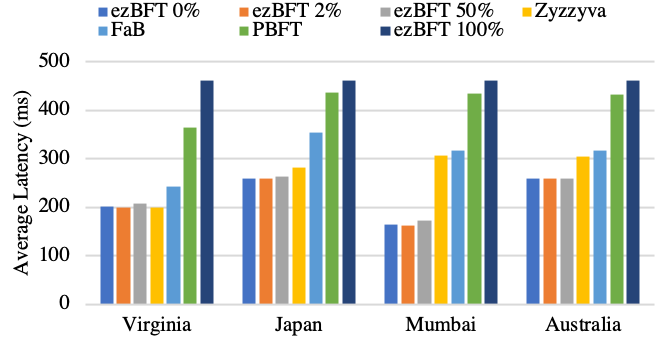}
	\caption{Average latencies for Experiment 1. All primaries are in US-East-1 region. The latency is shown per region as recorded by the clients in that region.}
	\label{fig:lat1}
\end{figure}

Figure~\ref{fig:lat1} shows the average latency (in milliseconds)  observed by the clients located in their respective regions (shown on x-axis) for each of the four protocols. For \thesystem, the latency was measured at different contention levels: 0\%, 2\%, 50\%, and 100\%; the suffix in the legend indicates the contention. Among primary-based protocols, PBFT suffers the highest latency, because it takes five communication steps to deliver a request. FaB performs better than PBFT with four communication steps, but Zyzzyva performs the best among primary-based protocols using only three communication steps. Overall, \thesystem performs as good as or better than Zyzzyva, for up to 50\% contention. In the US-East-1 region, both Zyzzyva and \thesystem have about the same latency because they have the same number of communication steps and their primaries are located in the same region. However, for the remaining regions, Zyzzva clients must forward their requests to US-East-1, while \thesystem clients simply send their requests to their local replica, which orders them. At 100\% contention, five communication steps required for total-order increases \thesystem's latency close to that of PBFT's. 

\subsubsection*{Experiment 2}
\begin{figure}[htbp]
	\centering
	\begin{subfigure}[b]{0.4\textwidth}
		\includegraphics[width=\textwidth]{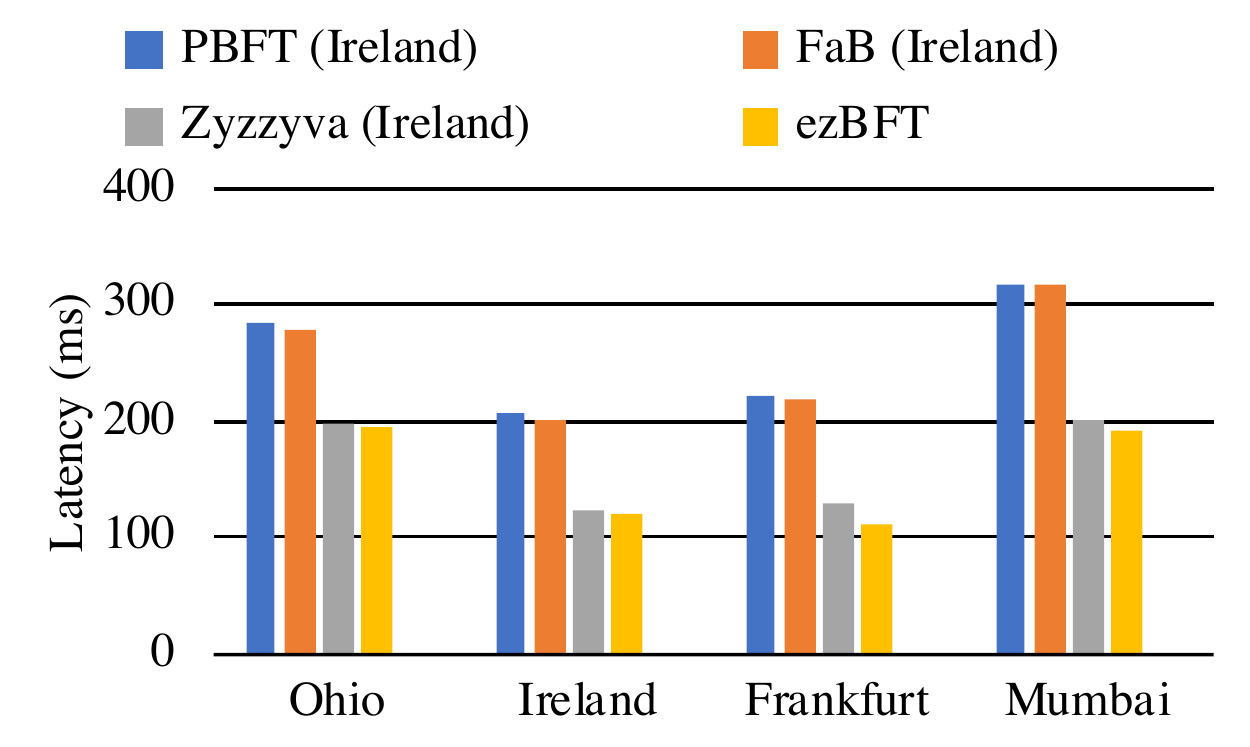}
		\caption{Average latencies for Experiment 2.}
		\label{fig:lat2a}
	\end{subfigure}
	\begin{subfigure}[b]{0.4\textwidth}
		\includegraphics[width=\textwidth]{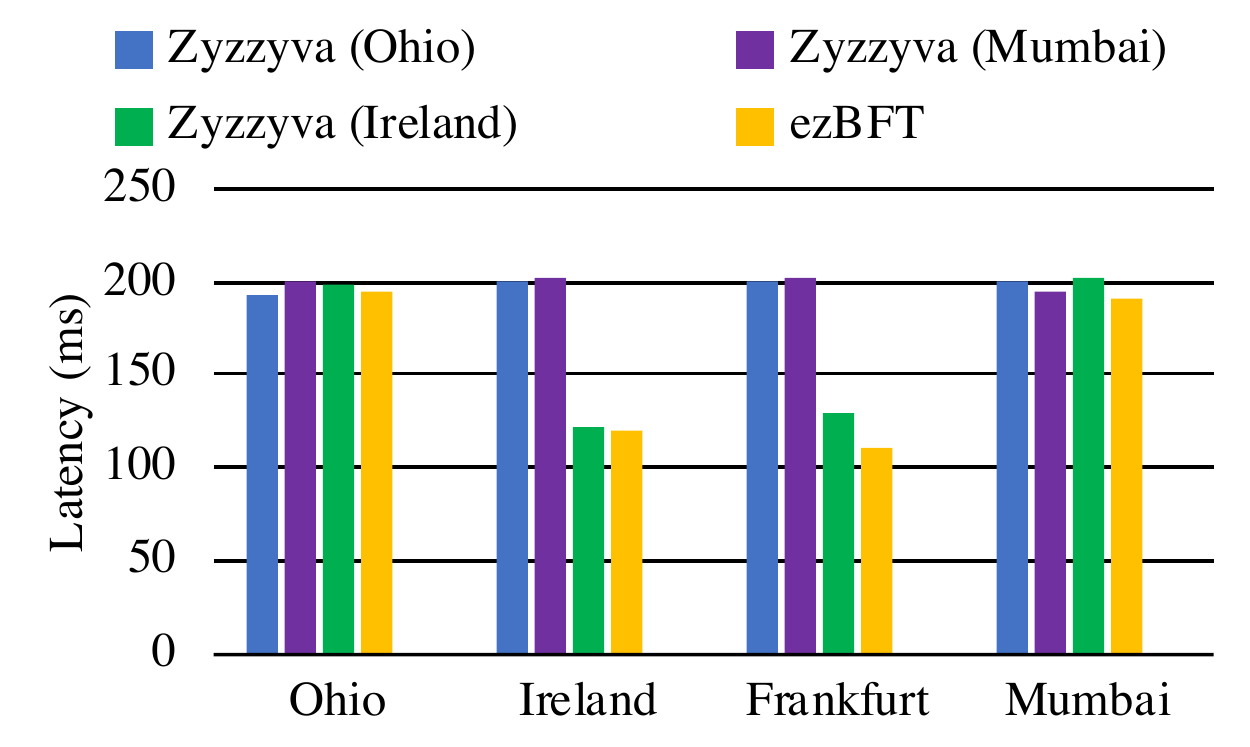}
		\caption{Average latencies for Experiment 2. Zyzzva's  primary is at different regions. }
		\label{fig:lat2b}
	\end{subfigure}
	\caption{Best case for Zyzzyva and the effect of moving primary to different regions. Experiments reveal  \thesystem's effectiveness. Legend entries show primary's  location in parenthesis.}
	\label{fig:lat2}
\end{figure}

To better understand Zyzzyva's best and worst-case performances and how they fare against \thesystem, we identified another set of AWS regions: US-East-2 (Ohio), Ireland, Frankfurt, and India. 
This experiment was run similar to that of Figure~\ref{fig:lat1}. 
The primary was placed in Ireland.
The results are shown in Figure~\ref{fig:lat2}.
Figure~\ref{fig:lat2a} shows the average latencies as observed by the clients in each deployed region for each of the four protocols. The choice of Ireland as the primary region represents the best case for Zyzzyva. Hence, \thesystem performs very similar to Zyzzyva.

In Experiment 1, 
the regions mostly had non-overlapping paths between them, and thus the first communication step of sending the request to the leader can be seen clearly (notice Mumbai in Figure~\ref{fig:lat1}. On the other hand, in Experiment 2, 
connections between the regions have overlapping paths. For example, sending a request from Ohio to Mumbai for \thesystem will take about the same time as sending a request from Ohio to Mumbai via the primary at Ireland for Zyzzyva.  

Figure~\ref{fig:lat2b} shows the effect of moving the primary to different regions. We disregard PBFT and FaB in this case, as their performance do not improve. For Zyzzyva, moving the primary  to US-East-2 or India substantially increases its overall latency. In such cases, \thesystem's latency is up to 45\% lower than Zyzzyva's.
This data-point is particularly important as it reveals how the primary's placement affects the latency. 

To curb the negative effects of byzantine primary replicas, in~\cite{castro2002practical}, the authors propose to frequently move the primary (this strategy is adopted by  other protocols including Zyzzyva). From Figure~\ref{fig:lat2b}, we can extrapolate that such frequent movements can negatively impact latencies over time. Given these challenges, we argue that \thesystem's leaderless nature is a better fit for geo-scale deployments. 

\subsection{Client Scalability}

\begin{figure}[htbp]
	\centering
	\includegraphics[width=\linewidth]{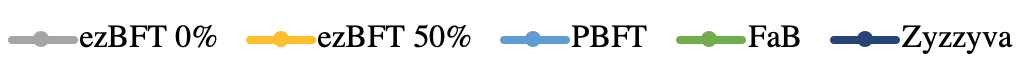}
	\includegraphics[width=\linewidth]{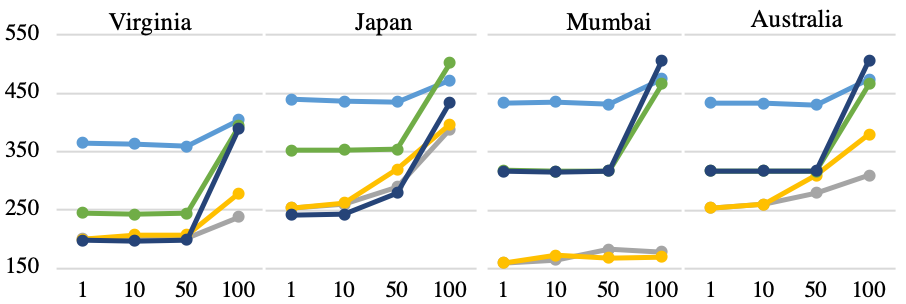}
	\caption{Latency per location while varying the number of connected clients (1 -- 100) per region.}
	\label{fig:latclient}
\end{figure}

Another important aspect of \thesystem is its ability to maintain low  client-side latency even as the number of connected clients increases. For this experiment, we deployed the protocols in Virginia, Japan, Mumbai, and Australia, and measured client-side latency per region by varying the number of connected clients. Figure~\ref{fig:latclient} shows the results. Notice that as Zyzzyva approaches 100 connected clients per region, it suffers from an exponential increase in latency. However, \thesystem, even at 50\% contention, is able to scale better with the number of clients. Particularly,  in Mumbai, \thesystem maintains a stable latency even at 100 clients per region, while Zyzzyva's latency shoots up.

\subsection{Server-side Throughput}

We also measured the peak throughput of the protocols. For this experiment, we deployed the protocols in five AWS regions: US-East-1 (Virginia), India, Australia, and Japan. We co-located ten clients with the primary replica in US-East-1. Unlike the experiments so far, here the clients send requests in an open-loop, meaning that they continuously and asynchronously send requests before receiving replies.  The requests consists of an 8-byte key and a 16-byte value.

\begin{figure}[htbp]
	\center
	\includegraphics[width=.48\textwidth]{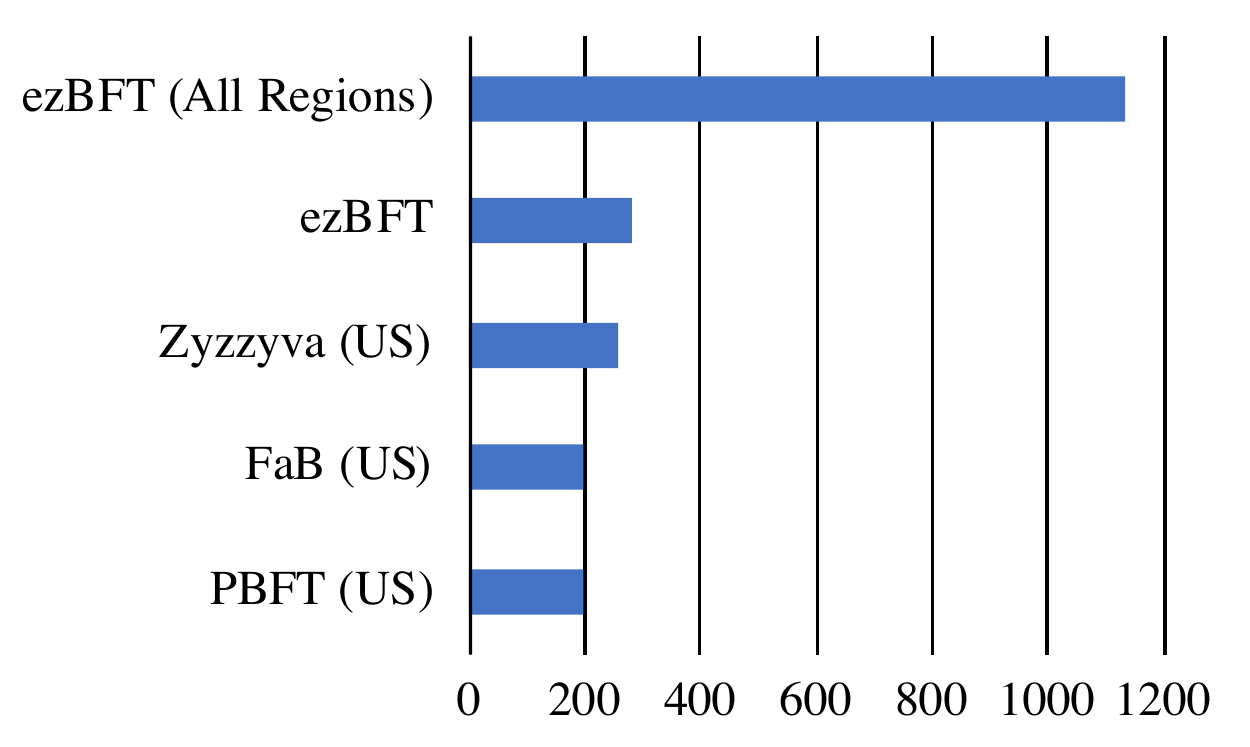}
	\caption{Throughput of \thesystem and competitor BFT protocols.}
	\label{fig:tps}
\end{figure}

Figure~\ref{fig:tps} shows the results. For \thesystem, we carried out two experiments: i) clients are placed only at US-East-1 (labelled \thesystem in the figure); and ii) clients are placed at every region (labelled ``\thesystem (All Regions)'' in the figure). Due to the leaderless characteristic, each of the replicas can feed requests into the system, increasing the overall throughput. The contention was set to 0\%, and no batching was done.

Observe that when clients are placed only at US-East-1, \thesystem performs at par or slightly better than others. On the other hand, when clients are placed in all the other regions, which does not yield any benefit for other protocols, \thesystem's throughput increases by as much as four times, as all \thesystem replicas are able to process and deliver requests concurrently.


%% file: relwork.tex
\section{Related Work}
\label{sec:relwork}


BFT consensus was first introduced in~\cite{lamport1982byzantine}. However, PBFT~\cite{castro1999practical, castro2002practical} was the first protocol to provide a practical implementation of BFT consensus in the context of state machine replication. PBFT  solved consensus in three communication steps (excluding two steps for client communication) using $3f + 1$ nodes and requiring responses from at least $2f+1$ nodes. $f$ is the maximum number of byzantine faulty nodes that the system can tolerate and make progress.

FaB~\cite{martin2006fast} reduced the number of communication steps required to reach agreement in the common case to two  steps, called two-step consensus (excluding two steps for client communication). However, node and quorum requirements are substantially larger with $5f+1$ and $4f+1$ nodes, respectively, to tolerate $f$ faults and still achieve two-step consensus. FaB falls back to a slower path when $4f+1$ responses cannot be acquired, and requires an extra communication step and at least $3f+1$ nodes to reach agreement. 

FaB was the first BFT protocol to not require any digital signatures in the common case. Parameterized-FaB~\cite{martin2006fast}
requires $3f + 2t + 1$ nodes, where $f$ is the number of tolerated faults and preserve safety and $t$ is the number of tolerated faults and solve two-step consensus. This minimized the node requirements in the common case to $3f+1$ (by setting $t=0$), but the quorum requirement is still more than that of PBFT.

The Q/U~\cite{Abd-El-Malek:2005:FBF:1095810.1095817} protocol was the first to achieve consensus in two communications steps when there are no faults and update requests do not concurrently access the same object. Q/U defines a simplified version of conflicts. Requests are  classified as either reads or writes. Reads do not conflict with reads, while write conflicts with reads and writes. This is more restrictive than the commutative property used by \thesystem. In \thesystem, for instance, mutative operations (such as incrementing a variable) are commutative.

HQ~\cite{cowling2006hq} is similar to PBFT  with a special optimization to execute read-only requests in two communication steps and update requests in four communication steps under no conflicts. HQ's definition of conflict is the same as Q/U's.

Zyzzyva~\cite{kotla2007zyzzyva, kotla2009zyzzyva} uses $3f+1$ nodes to solve consensus in three steps (including client communication), requiring a quorum of $3f+1$ responses. The protocol tolerates $f$ faults, so it takes an additional two  steps when nodes are faulty. Zyzzyva uses the  minimum number of nodes, communication steps, and one-to-one messages to achieve fast consensus. It is cheaper than the  aforementioned protocols, but also more complex. Zyzzyva's  performance boost is due to speculative execution, active participation of the clients in the agreement process, and tolerating temporary inconsistencies among replicas. 

\thesystem has the same node and quorum requirements as well as the number of communication steps as Zyzzyva. However, by minimizing the latency of the first communication step and alleviating the specialized role of the primary, \thesystem reduces the request processing latency.

Aliph~\cite{guerraoui2010next} builds a BFT protocol by composing three different sub-protocols, each handling a  specific system factor such as contention, slow links, and byzantine faults. Under zero contention, the sub-protocol \emph{Quorum} can deliver agreement in two  steps with $3f+1$ nodes by allowing clients to send the requests directly to the nodes. However, as contention or link latency increases, or as faults occur, Aliph switches to the sub-protocol \emph{Chain} whose additional steps is equal to the number of nodes in the system, or to the sub-protocol \emph{Backup} which takes at least three  steps. 
Although the idea of composing simpler protocols is appealing in terms of reduced design and implementation complexities, the performance penalty is simply too high, especially in geo-scale settings.

In contrast, \thesystem exploits the trade-off between the slow and fast path  steps. \thesystem uses three steps compared to Aliph's two steps in the common case, and in return, provides slow path in only two extra communication steps unlike Aliph. Moreover, \thesystem's leaderless approach reduces the latency of the first communication step to near zero, yielding client-side latency  comparable to Aliph's.

EBAWA~\cite{5634304} uses the spinning primary approach~\cite{veronese2009spin} to minimize the client-side latency in geo-scale deployments. However, a byzantine replica can delay its commands without detection reducing the overall server-side throughput. \thesystem's dependency collection mechanism enables correct replicas to only depend on commands that arrive in time, and execute without waiting otherwise.


Table~\ref{tab:comparison} summarizes the comparison of existing work with \thesystem. Note that \thesystem and Zyzzyva have the same best-case communication steps. However, for \thesystem, the latency for the first-step communication is minuscule (tending towards zero) when compared to Zyzzyva's first-step latency.

\begin{table}[t]
	\centering
	\caption{Comparison of existing BFT protocols and \thesystem.}
	\begin{tabular}{|l||c|c|c|c|}
		\hline
		\bf Protocol                                                   &    \bf PBFT     &                \bf Zyzzyva                 &                  \bf Aliph                  &                  \bf ezBFT                  \\ \hline\hline
		Resilience                                                     &    $f < n/3$    &                  $f < n/3$                  &                  $f < n/3$                  &                  $f < n/3$                  \\ \hline
		\makecell[l]{Best-case \\ comm. steps}                         &        5        &                      3                      &                      2                      &                      3                      \\ \hline
		\makecell[l]{Best-case \\ comm. steps \\ in absence of \\ ...} & \makecell{Byz. \\ Slow links} & \makecell{Byz. \\ Slow links \\ Contention} & \makecell{Byz. \\ Slow links \\ Contention} & \makecell{Byz. \\ Slow links \\ Contention} \\ \hline
		Slow-path steps                                                &        -        &                      2                      &                    n + 3                    &                      2                      \\ \hline
		Leader                                                         &     Single      &                   Single                    &                   Single                    &                 Leaderless                  \\ \hline
	\end{tabular}
	\label{tab:comparison}
\end{table}

Leaderless and multi-leader protocols~\cite{caesar, peluso2016making, moraru2013there, DBLP:conf/osdi/MaoJM08} have been proposed for the CFT model. Among these, EPaxos~\cite{moraru2013there} 
and Caesar~\cite{caesar} 
collect dependencies and find a total order among all the dependent requests. Both protocols work in two phases: a fast phase that is reached under no contention and an additional slow phase that is required under contention. In EPaxos, the collected dependencies form a graph, which is linearized before  executed. 

Caesar  
can deliver fast phase consensus even under non-trivial contention by having a replica wait until some conditions are satisfied before replying to the primary. Such wait conditions are harmful in BFT protocols, because a malicious replica can use this as an opportunity to cease progress.

In $M^2$Paxos~\cite{peluso2016making}, a replica can order a request if it \emph{owns} the object that the request accesses. Otherwise, it forwards the request to the right owner or acquires ownership. Acquiring ownership means becoming the primary for some subset of objects, and in CFT-based protocols, any replica can propose to be a owner of any subset of objects at any point in time. However, in BFT-based protocols, electing a primary is a more involved process requiring consent from other replicas. In addition, view numbers are pre-assigned to replicas; therefore, randomly choosing primaries is not a trivial process.


%% file: conclusion.tex
\section{Conclusions}
\label{sec:conclusion}

State-of-the-art BFT protocols are not able to provide optimal request processing latencies in geo-scale deployments -- an increasingly ubiquitous scenario for many distributed applications, particularly blockchain-based applications. 

We presented \thesystem, a leaderless BFT protocol that provides three-step consensus in the common case, while essentially nullifying the latency of the first communication step. \thesystem provides the classic properties of BFT protocols including nontriviality, consistency, stability, and liveness. 
Our experimental evaluation reveals that \thesystem reduces latency by up to 40\% compared to Zyzzyva.